\documentclass[11pt,a4paper]{article}
\usepackage{jheppub}

\usepackage{amsmath}
\usepackage{amsfonts}
\usepackage{amsthm}
\usepackage{slashed}
\usepackage{graphicx}
\usepackage{mathrsfs,bbm}
\usepackage{subfig}
\usepackage{booktabs}
\usepackage[numbers,sort&compress]{natbib}

\makeatletter
\g@addto@macro\bfseries{\boldmath}
\makeatother

\theoremstyle{remark}
\newtheorem*{remark}{Remark}

\newcommand{\dimeps}{\varepsilon}
\newcommand{\td}{\mathrm{d}}
\newcommand{\iu}{\mathrm{i}}
\newcommand{\ieps}{\iu\epsilon}
\newcommand{\FI}{\mathcal{I}}
\newcommand{\TOP}[1][top]{\mathsf{#1}}
\newcommand{\PL}{\TOP[PL]}
\newcommand{\NP}{\TOP[NP]}
\newcommand{\FIPL}[1]{\FI_{\PL}(#1)}
\newcommand{\FINP}[1]{\FI_{\NP}(#1)}
\newcommand{\F}{\mathcal{F}}
\newcommand{\U}{\mathcal{U}}
\newcommand{\sdc}{\omega}
\newcommand{\Transpose}{\intercal}

\newcommand{\HyperInt}{\href{http://bitbucket.org/PanzerErik/hyperint/}{\texttt{\textup{HyperInt}}}}
\newcommand{\GiNaC}{\texttt{GiNaC}}
\newcommand{\Li}[1]{\operatorname{Li}_{#1}}
\newcommand{\abs}[1]{\left| #1 \right|}

\title{Two-loop mixed QCD-EW corrections to \boldmath $gg \to Hg$}

\author[a,b]{Marco Bonetti} 
\author[c]{Erik Panzer}
\author[d,e]{Vladimir A. Smirnov}
\author[f]{Lorenzo Tancredi}

\affiliation[a]{Institute for Theoretical Particle Physics, KIT, \\Wolfgang-Gaede-Strasse 1, D-76128 Karlsruhe, Germany}
\affiliation[b]{Institute for Theoretical Particle Physics and Cosmology, RWTH Aachen University, \\Sommerfeldstrasse 16, D-52056 Aachen, Germany} 
\affiliation[c]{All Souls College, University of Oxford, \\OX1 4AL, Oxford, U.K.}
\affiliation[d]{Skobeltsyn Institute of Nuclear Physics of Moscow State University, \\119991 Moscow, Russia}
\affiliation[e]{Moscow Center for Fundamental and Applied Mathematics, \\119991 Moscow, Russia}
\affiliation[f]{Rudolf Peierls Centre for Theoretical Physics, University of Oxford, Clarendon Laboratory, \\Parks Road, OX1 3PU, Oxford, U.K.}

\emailAdd{bonetti@physik.rwth-aachen.de}
\emailAdd{erik.panzer@maths.ox.ac.uk}
\emailAdd{smirnov@theory.sinp.msu.ru}
\emailAdd{lorenzo.tancredi@physics.ox.ac.uk}

\abstract{We compute the two-loop mixed QCD-Electroweak (QCD-EW) corrections to the production of a 
Higgs boson and a gluon in gluon fusion through a loop of light quarks. The relevant four-point functions with internal massive propagators are expressed as multiple polylogarithms with algebraic arguments. We perform the calculation by integration over Feynman parameters and, independently, by the method of differential equations.
We compute the two independent helicity amplitudes for the process and we find that they are both finite. 
Moreover, we observe a weight drop when all gluons have the same helicity.
We also provide a simplified expression for the all-plus helicity amplitude, which is optimised
for fast and reliable numerical evaluation in the physical region.
}

\preprint{\begin{minipage}[t]{8cm}\begin{flushright} P3H-20-036, TTK-20-22, \\TTP20-027, OUTP-20-08P\\
      \end{flushright}\end{minipage}}

\begin{document}

\maketitle

\catcode`\@=11
\font\manfnt=manfnt
\def\Watchout{\@ifnextchar [{\W@tchout}{\W@tchout[1]}}
\def\W@tchout[#1]{{\manfnt\@tempcnta#1\relax%
  \@whilenum\@tempcnta>\z@\do{%
    \char"7F\hskip 0.3em\advance\@tempcnta\m@ne}}}
\let\foo\W@tchout
\def\dubious{\@ifnextchar[{\@dubious}{\@dubious[1]}}
\let\enddubious\endlist
\def\@dubious[#1]{%
  \setbox\@tempboxa\hbox{\@W@tchout#1}
  \@tempdima\wd\@tempboxa
  \list{}{\leftmargin\@tempdima}\item[\hbox to 0pt{\hss\@W@tchout#1}]}
\def\@W@tchout#1{\W@tchout[#1]}
\catcode`\@=12


\section{Introduction}
\label{sec:intro}

The discovery of the Higgs boson at the LHC~\cite{Aad:2012tfa,Chatrchyan:2012ufa} has marked a turning point
in the exploration of the Standard Model of particle physics. Not only is the Higgs boson
the only elementary scalar particle in the Standard Model, but it is also
related to the Electro-Weak (EW) symmetry breaking mechanism, which is believed to be 
responsible for the observed values of the masses of all elementary particles.
For this reason, the discovery of the Higgs boson and the measurement of its properties allow us to investigate 
the least studied aspects of the Standard Model.

Theory has to support this program by providing precise predictions for the Higgs production cross sections.
A special role here is played by the process $gg \to H+X$, which represents 
by far the largest Higgs production channel at the LHC.
The main contribution to this channel is provided by those Feynman diagrams where the Higgs boson couples to the gluons through a top quark loop. 
Given its importance, this process started receiving 
attention already many decades ago, and  today it is known
 up to next-to-leading order (NLO) in QCD~\cite{Georgi:1977gs,Graudenz:1992pv,Spira:1995rr,Aglietti:2006tp}.
While in those papers it was shown that the NLO corrections can be as large as $\mathcal{O}(100\%)$, 
an NNLO calculation with full top-mass 
dependence remains prohibitively complicated still today, primarily due to the complexity of the relevant three-loop massive 
scattering amplitudes. We note that recently 
the first numerical results for the relevant three-loop contributions have been obtained in~\cite{Czakon:2020vql}.

A surprisingly reliable way to estimate higher-order QCD corrections to the $gg \to H+X$ cross-section is
provided by studying this process in the limit of infinite top quark mass, where
the interaction between gluons and the Higgs boson is shrunk to a point-like effective vertex. Calculations in this limit
are substantially simpler than in the full theory, which made it possible to  push the perturbative expansion 
to NNLO~\cite{Harlander:2002wh,Anastasiou:2002yz,Ravindran:2003um} and more recently 
up to N$^3$LO~\cite{Anastasiou:2015vya,Mistlberger:2018etf} in perturbative QCD.
The N$^3$LO corrections amount to around $\sim 5\%$ of the total cross section and show a very 
good convergence of the QCD perturbative series, 
reducing the scale-uncertainty to $\sim 2\%$~\cite{Anastasiou:2016cez}.

At this level of precision, other contributions to Higgs production cannot be neglected anymore. 
One such contribution is given by the class of two-loop diagrams where the gluons couple to a 
loop of massless quarks, followed by two massive electroweak vector bosons, which finally fuse into a Higgs boson. 
Clearly, at this perturbative order, also other classes of diagrams contribute, 
where the Higgs boson couples directly to top-quarks.
These contributions are particularly difficult to compute because of the large number of internal masses, 
but we expect their size to be less than $\sim 15\%$ compared to those induced by massless quarks, 
at least close to threshold production~\cite{Degrassi:2004mx}. For this reason, in what follows we will
limit ourselves to consider massless quarks only.
These EW corrections have been computed at LO and have been shown to contribute up to $\sim 5\%$ to the 
gluon-fusion cross section~\cite{Aglietti:2004nj,Actis:2008ug}. Given that  NLO QCD corrections to 
gluon induced processes are typically large, it becomes  very important to have a reliable estimate of the QCD corrections
to this class of diagrams. Unfortunately, the calculation of these mixed NLO QCD-EW corrections is highly non-trivial, 
as it involves virtual three-loop three-point
diagrams and real-emission two-loop four-point diagrams with massive internal propagators.
While the former have been recently computed with full dependence on the Higgs and on the vector-boson masses~\cite{Bonetti:2017ovy},
the computation of the latter has remained an outstanding challenge and, before this paper, only the relevant
planar master integrals were known analytically~\cite{Becchetti:2018xsk}.

To overcome the complexity of the full calculation, different approximations have been employed to estimate the impact of these 
corrections. In particular, the mixed QCD-EW corrections have first been computed in the unphysical limit 
$m_V \gg m_H$~\cite{Anastasiou:2008tj}, where
they effectively reduce to a Wilson coefficient for the operator $\mathcal{O} = H G^{\mu \nu} G_{\mu \nu}$ and one 
therefore expects a  K-factor similar to the one in the NLO QCD heavy-top approximation.
This a priori unphysical approximation has recently been improved in~\cite{Bonetti:2018ukf}, where the exact results for the virtual amplitudes
computed in Ref.~\cite{Bonetti:2017ovy} have been combined with the real radiation computed in the soft-gluon approximation.
The soft-gluon approximation is known
to work relatively well for Higgs boson production~\cite{Catani:2001ic,deFlorian:2012za,Ball:2013bra} and
the calculation showed that accounting for finite vector boson masses in the virtual corrections provides consistent results with the 
Wilson-coefficient approximation employed in~\cite{Anastasiou:2008tj}.

The soft-gluon approximation amounts to the factorisation of the QCD and EW corrections in the real corrections.
One could therefore wonder if a breaking 
of this factorisation in the real-radiation pattern could modify the K-factor in a non-trivial way.
To estimate how good this approximation is, the mixed QCD-EW corrections  have also 
been considered
in the limit $m_V \to 0$~\cite{Anastasiou:2018adr}. This study confirmed that for small vector boson masses, 
the non-factorisable QCD-EW corrections remain negligible.
Clearly, this does not exclude the possibility that keeping full dependence on the masses of the 
electroweak vector bosons could induce non-negligible modifications to the NLO corrections. 
It remains therefore  very desirable 
 to compute exactly 
the missing two-loop QCD-EW real amplitudes  in order to provide a definite answer to this question.
As hinted to above, this calculation is also interesting on a formal level,
in particular due to the large number of scales and to the vector boson masses
in the internal propagators,
which translate into an involved analytic structure of the corresponding Feynman integrals.

Specifically, we find that 
the relevant Feynman integrals can be expressed in terms of multiple polylogarithms~\cite{Nielsen,Goncharov:1998kja,Remiddi:1999ew,Goncharov:2001iea} 
with algebraic arguments, involving multiple square roots. While the standard approach to compute such integrals
would go through the derivation and solution of differential equations in canonical form~\cite{Kotikov:1990kg,Remiddi:1997ny,Gehrmann:1999as,Henn:2013pwa},  
the complexity of the alphabet makes this strategy extremely cumbersome in practice.  
Interestingly, though, we find that all relevant integrals can be computed by
integrating over Feynman parameters using the algorithms described in~\cite{Panzer:2014caa,Brown:2008um}.
The results thus obtained turn out  to be very compact, but not extremely efficient for the numerical evaluation of the amplitude
in Minkowski kinematics. This provides us with the ground to discuss a general strategy for their simplification
and to present alternative results for the amplitude which are of more direct use for phase-space integration.

Finally, we stress that in this paper we only consider  the two-loop real scattering amplitudes for the NLO QCD-EW corrections to $gg \to Hg$.
While we do not expect them to  constitute any additional complexity, we do not consider quark-initiated partonic channels, 
whose contribution has been shown to be negligible at this precision~\cite{Keung:2009bs}.

The rest of the paper is organised as follows. In section~\ref{sec:notation} we give our notation and describe how to compute the helicity amplitudes
for $gg \to Hg$ by decomposing the amplitude into form factors with the help of $d$-dimensional projection operators.
After describing the reduction to master integrals and our choice of basis, we explain in sections~\ref{sec:intsDEQs} and~\ref{sec:finite-basis} the calculation of the master integrals with two independent approaches, i.e.\ using differential equations and by parametric integration, respectively.
We discuss our final result for the helicity amplitudes in section~\ref{sec:results} and finally conclude in section~\ref{sec:conc}.


\section{The scattering amplitudes}
\label{sec:notation}

\begin{figure}[h!]
\centering
       \subfloat[]{{\includegraphics[width=5cm]{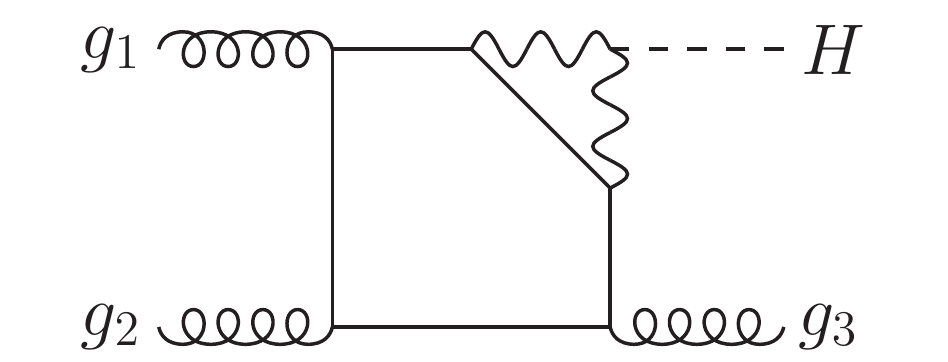} }}%
  \qquad
  \subfloat[]{{\includegraphics[width=5cm]{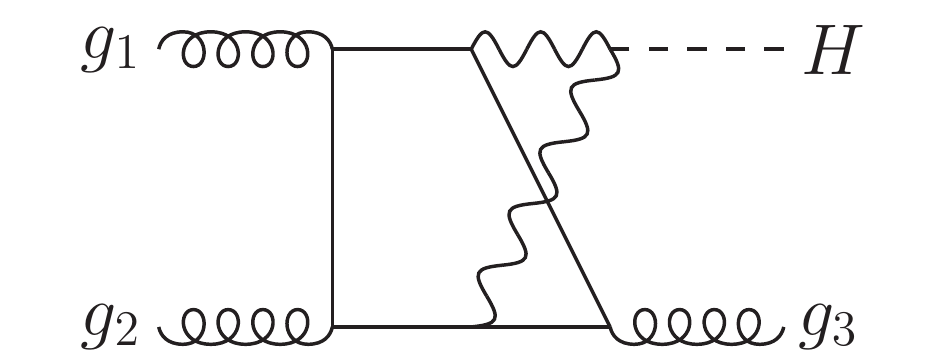} }}%
    \caption{Representative planar (a) and non-planar (b) Feynman diagrams for the LO mixed QCD-EW corrections to $gg \to Hg$. The internal wavy lines
    represent the massive vector bosons.}%
    \label{fig:diagrams}%
\end{figure}

We are interested in computing the two-loop mixed QCD-EW corrections to the production of a Higgs boson and a gluon in gluon
fusion at the LHC. We begin by considering the process in the decay kinematics
\begin{equation}
H(p_4) \to g(p_1) + g(p_2) + g(p_3) 
\end{equation}
where the Higgs couples to the gluons through a pair of massive vector bosons V = $Z$,$W^\pm$ and a 
massless quark loop, see Figure~\ref{fig:diagrams}. 

The scattering amplitude for this process depends on the three Mandelstam variables
\begin{equation}
s = (p_1 + p_2)^2\,, \quad t = (p_1 + p_3)^2\,, \quad u = (p_2 + p_3)^2\,, \quad \mbox{with} \quad s+t+u = m_h^2\,,
\end{equation}
and on the mass of the vector boson that mediates the interaction with the Higgs and which we will generically denote 
as $m_V$. We use $m_h$ to indicate the Higgs mass.
Since the QCD-EW contributions to $gg\to Hg$ start at two-loop order, the amplitudes computed in this
paper are finite, as long as all external gluons are fully resolved.

In order to perform the computation, we begin by decomposing the scattering amplitude for $H \to ggg$ 
into Lorentz- and gauge-invariant 
tensor structures. We extract the dependence on the $SU(3)$ color structure $f^{abc}$ and write 
\begin{equation}
A(p_1,p_2,p_3) = f^{a_1 a_2 a_3} \epsilon_1^\mu \epsilon_2^\nu \epsilon_3^\rho \, \mathcal{A}_{\mu \nu \rho}(s,t,u,m_V^2) \label{eq:tensorA}
\end{equation}
where, for each $j=1,2,3$, $\epsilon_j$ is the polarisation vector of the gluon of momentum $p_j$, while $a_j$ is its color label.
$\mathcal{A}_{\mu \nu \rho}(s,t,u,m_V^2)$ must be a rank-3 tensor under 
Lorentz transformations and, imposing gauge invariance for 
each of the external gluons, it can be written as a linear combination of four independent form factors. 
Following the conventions introduced in~\cite{Melnikov:2016qoc},
we require that the gluons are transverse and make a cyclic choice for their gauge fixing condition
\begin{align}
\epsilon_j \cdot p_j = 0\,, \quad \epsilon_1 \cdot p_2 = \epsilon_2 \cdot p_3 = \epsilon_3 \cdot p_1 = 0\,. \label{eq:gauge}
\end{align}
With this choice, one easily finds~\cite{Melnikov:2016qoc}
\begin{align}
\mathcal{A}^{\mu \nu \rho}(s,t,u,m_V^2) &= F_1(s,t,u,m_V^2) g^{\mu \nu} p_2^\rho 
                                                                   + F_2(s,t,u,m_V^2) g^{\mu \rho} p_1^\nu \nonumber \\
                                                                &+ F_3(s,t,u,m_V^2) g^{\nu \rho} p_3^\mu 
                                                                   + F_4(s,t,u,m_V^2) p_3^\mu p_1^\nu p_2^\rho\,, \label{eq:ff}
\end{align}
where the $F_j(s,t,u,m_V^2)$ are Lorentz-invariant  form factors. We stress that in this decomposition no parity-violating 
terms appear. 
This can be justified by noticing that, if we only consider massless quarks, 
the axial contribution drops when summing over degenerate isospin doublets. Clearly,
this cancellation does not happen for the
third quark doublet, where the mass degeneration is broken and the contribution from
the bottom quark alone is not well defined without the corresponding top-induced diagrams. 
As it is common practice when working in the framework of massless QCD, 
we deal with this issue by not allowing bottom quarks to propagate in the diagrams where $W$ 
bosons are exchanged, but by keeping them in all other diagrams. 
The missing axial contributions
from these diagrams are expected to be suppressed.\footnote{Moreover, these contributions should be 
proportional to the color factor $d^{abc}$, which
drops in the cross-section when contracted with the leading tree-level amplitudes for the process $gg \to gH$ 
coming from the infinite top mass effective theory.}
With this, we can write for each form factor 
\begin{align}
	\label{eq:ffEW}
F_j(s,t,u,m_V^2) =  -\frac{(\alpha \, \alpha_s)^{3/2} m_W}{16 \pi \sin^3{\theta_W}} \,  C_V \, \left( \mathcal{F}^{(0)}_j(s,t,u,m_V^2) + \mathcal{O}(\alpha_s,\alpha) \right)\,, \quad j=1,\ldots,4, 
\end{align}
where 
\begin{align}
C_W = 4\,, \qquad C_Z = \frac{2}{\cos^4{\theta_W}} \left( \frac{5}{4} - \frac{7}{3} \sin^2{\theta_W} + \frac{22}{9} \sin^4{\theta_W} \right) \,,
\end{align}
and $\mathcal{O}(\alpha_s,\alpha)$ indicates higher order contributions both in the QCD and in the EW coupling. 
The full QCD-EW corrections can then  be obtained by summing the contributions with $V=Z$ or $W$.

The form factors $F_j$, or equivalently the $\mathcal{F}_j^{(0)}$, are not the objects that we are ultimately interested in. 
Indeed, often 
substantial simplifications occur when one combines the form factors to compute so-called helicity amplitudes.
For the case at hand, each gluon can have two different helicities for a total of eight different combinations.
By use of Bose symmetry, parity and charge conjugation, one can easily show that only two of them are independent.
We indicate the helicity of the gluon of momentum $p_j$ by $\lambda_j$ and  write for a generic helicity amplitude and for a given
vector boson $V$
\begin{equation}
\mathcal{A}_{\lambda_1 \lambda_2 \lambda_3} (s,t,u,m_V^2) = \epsilon_{1,\lambda_1}^\mu(p_1)  \epsilon_{2,\lambda_2}^\nu(p_2)  \epsilon_{3,\lambda_3}^\rho(p_3)
\mathcal{A}_{\mu \nu \rho}(s,t,u,m_V^2) 
\end{equation}
where $\mathcal{A}_{\mu \nu \rho}(s,t,u,m_V^2) $ was defined in eq.~\eqref{eq:tensorA}.
We proceed by choosing as two independent helicity amplitudes $\mathcal{A}_{++\pm} (s,t,u,m_V^2)$.
It is straightforward to find compact expressions for these amplitudes in terms of the form factors in eq.~\eqref{eq:ff} 
 using the spinor-helicity formalism, see~\cite{Dixon:1996wi} and references therein. 
 We choose for the polarisation vectors of the external gluons
\begin{align}
\epsilon_{j,+}^\mu(p_j) = \frac{\langle q_j | \gamma^\mu | j ]}{\sqrt{2} \langle q_j j \rangle}\,,
\qquad
\epsilon_{j,-}^\mu(p_j) = - \frac{[ q_j | \gamma^\mu | j \rangle }{\sqrt{2} [ q_j j ]}\,,
\end{align}
where $q_j$ is an arbitrary reference vector with $q_j^2 = 0$ and $q_j \cdot p_j \neq 0$. While in principle the vector $q_j$ can be chosen freely,
the conditions in eq.~\eqref{eq:gauge} force us to pick $q_1 = p_2$, $q_2 = p_3$ and $q_3 = p_1$.
With this, the two independent helicity amplitudes become
\begin{equation}\label{eq:helamplit}\begin{split}
&\mathcal{A}_{+++} (s,t,u,m_V^2) = \frac{m_h^2}{ \sqrt{2} \langle 12 \rangle \langle 23 \rangle \langle 31 \rangle} \Omega_{+++}(s,t,u,m_V^2)\,, \\
&\mathcal{A}_{++-} (s,t,u,m_V^2) = \frac{[12]^3}{ \sqrt{2} [13][23] m_h^2 }  \Omega_{++-}(s,t,u,m_V^2)\,,
\end{split}\end{equation}
where the $\Omega_{++\pm}$ are linear combinations of the original form factors
\begin{align}
\Omega_{+++} = \frac{s u}{m_h^2} \left( F_1 + \frac{t}{u} F_2 + \frac{t}{s} F_3 + \frac{t}{2} F_4 \right)\,, \qquad
\Omega_{++-} = \frac{m_h^2 u}{s} \left( F_1 + \frac{t}{2} F_4 \right)\,. \label{eq:helampl}
\end{align}
Similarly to eq.~\eqref{eq:ffEW}, we can explicitly extract the LO EW and QCD couplings from the amplitudes and write for the perturbative expansion
of the 
helicity coefficients
\begin{align}
\Omega_{++\pm} (s,t,u,m_V^2) = -\frac{(\alpha \, \alpha_s)^{3/2} m_W}{16 \pi \sin^3{\theta_W}} \,  C_V \, \left( \Omega^{(0)}_{++\pm}(s,t,u,m_V^2) + \mathcal{O}(\alpha_s,\alpha) \right)\,,
\end{align}
such that, again, the full QCD-EW contributions are obtained by summing the corresponding helicity amplitudes with $V=Z,W$.
 
\subsection{The evaluation of the helicity amplitudes}
\label{sec:amp-evaluation}
The helicity amplitudes in eq.~\eqref{eq:helampl} receive contribution from $21$ different two-loop Feynman diagrams, see Figure~\ref{fig:diagrams} for
two representative ones. The contribution of each of these diagrams to the helicity coefficients can be computed by deriving $d$-dimensional
projector operators. The standard approach consists of introducing 4 independent projectors which single out the contribution to each of the
form factors defined in eq.~\eqref{eq:ff} 
\begin{align}
\sum_{pol}\, P_j^{\mu \nu \rho}\, 
(\epsilon_1^{\mu})^* \epsilon_1^{\mu_1}\;
(\epsilon_2^{\nu})^* \epsilon_2^{\nu_1}\;
(\epsilon_3^{\rho})^* \epsilon_3^{\rho_1}\; {\cal A}_{\mu_1 \nu_1 \rho_1}(s,t,u, m_V^2)  = F_j(s,t,u,m_V^2)\,,
\label{eq:projdef}
\end{align}
where, for consistency with eq.~\eqref{eq:gauge},  we must use
\begin{align}
&\sum_{pol} \left( \epsilon_1^{\mu}(p_1) \right)^* \epsilon_1^{\nu}(p_1) = 
- g^{\mu \nu} + \frac{p_1^\mu p_2^\nu + p_1^\nu p_2^\mu}{p_1 \cdot p_2}\,, \label{eq:polsums1}\\
&\sum_{pol} \left( \epsilon_2^{\mu}(p_2) \right)^* \epsilon_2^{\nu}(p_2) = 
- g^{\mu \nu} + \frac{p_2^\mu p_3^\nu + p_2^\nu p_3^\mu}{p_2 \cdot p_3}\,,  \label{eq:polsums2}\\
&\sum_{pol} \left( \epsilon_3^{\mu}(p_3) \right)^* \epsilon_3^{\nu}(p_3) = 
- g^{\mu \nu} + \frac{p_1^\mu p_3^\nu + p_1^\nu p_3^\mu}{p_1 \cdot p_3} \,. \label{eq:polsums3}
\end{align}
We stress at this point  that all Lorentz indices in eq.~(\ref{eq:projdef}) 
have to be understood  as  $d$-dimensional. 
Each projector can be decomposed in terms of the same tensor structures as in eq.~\eqref{eq:ff} as follows
\begin{align}
P_j^{\mu \nu \rho} &= \frac{1}{d-3} 
\left [ c_1^{(j)}\, g^{\mu \nu}\, p_2^\rho
 + c_2^{(j)}\, g^{\mu \rho}\, p_1^\nu + c_3^{(j)}\, g^{\nu \rho}\, p_3^\mu
 + c_4^{(j)}\, p_3^\mu p_1^\nu p_2^\rho\, \right], 
\end{align}
where $ j \in \{1,2,3,4 \}$. By imposing that 
eq.~\eqref{eq:projdef} is satisfied we find
\begin{equation}\begin{aligned}
c_1^{(1)} &= \frac{t}{s\,u}\,,&
c_2^{(1)} &= 0\,,&
c_3^{(1)} &= 0\,,&
c_4^{(1)} &= -\frac{1}{s\,u}\,, 
\\
c_1^{(2)} &= 0\,,&
c_2^{(2)} &= \frac{u}{s\,t}\,,&
c_3^{(2)} &= 0\,,&
c_4^{(2)} &= -\frac{1}{s\,t}\,,
\\
c_1^{(3)} &= 0\,,&
c_2^{(3)} &= 0\,,&
c_3^{(3)} &= \frac{s}{t\,u}\,,&
c_4^{(3)} &= -\frac{1}{t\,u}\,,
\\
c_1^{(4)} &= -\frac{1}{s\,u}\,,\quad&
c_2^{(4)} &= -\frac{1}{s\,t}\,,\quad&
c_3^{(4)} &= -\frac{1}{t\,u}\,,\quad&
c_4^{(4)} &=  \frac{d}{s\,t\,u}\,.
\end{aligned}\end{equation}

We can either use these projectors to evaluate the four form factors independently, or we can use them, together with the definition 
of the helicity coefficients in terms of form factors in eq.~\eqref{eq:helampl},
 in order to derive new helicity-projectors~\cite{Peraro:2019cjj} that directly project on the physical helicity amplitudes
 \begin{equation}
\begin{split}
 P_{+++}^{\mu \nu \rho} &= \frac{1}{2 m_h^2(d-3)} \Big[ t\, g^{\mu \nu}\, p_2^\rho 
 +  u\, g^{\mu \rho}\, p_1^\nu   
 + s\, g^{\nu \rho}\, p_3^\mu 
 + (d-6)\, p_3^\mu p_1^\nu p_2^\rho \Big] \,, \\
 P_{++-}^{\mu \nu \rho}  &= \frac{m_h^2}{2 s^2(d-3)}\, \Big[ t\, g^{\mu \nu}\, p_2^\rho 
 - u\, g^{\mu \rho}\, p_1^\nu   
 - s \, g^{\nu \rho}\, p_3^\mu 
 + (d-2)\, p_3^\mu p_1^\nu p_2^\rho \Big]\,, \label{eq:projOmegas}
 \end{split}
 \end{equation}
such that
\begin{align}
\sum_{pol}\, P_{++\pm}^{\mu \nu \rho}\, 
(\epsilon_1^{\mu})^* \epsilon_1^{\mu_1}\;
(\epsilon_2^{\nu})^* \epsilon_2^{\nu_1}\;
(\epsilon_3^{\rho})^* \epsilon_3^{\rho_1}\; {\cal A}_{\mu_1 \nu_1 \rho_1}(s,t,u, m_V^2)  = \Omega_{++\pm}(s,t,u,m_V^2)\,.
\end{align}
Since the helicity amplitudes are the physical objects that we will be ultimately interested in, we prefer to 
follow this second approach.

In practice, we generate all relevant two-loop diagrams using \texttt{QGRAF}~\cite{Nogueira:1991ex} and we use \texttt{FORM}~\cite{Vermaseren:2000nd} 
to apply the projectors in eq.~\eqref{eq:projOmegas} and  
write them as linear combinations of scalar two-loop Feynman integrals.
We find that all diagrams can be mapped on Feynman integrals of two integral families, 
one planar ($\PL$) and one non-planar ($\NP$), up to crossings of the
external legs. We define these two families as follows:
\begin{equation}
	\FI_{\TOP}(a_1,a_2,\ldots,a_8,a_9)
= \int 
\frac{\mathfrak{D}^dk\, \mathfrak{D}^dl}{D_1^{a_1} D_2^{a_2}  D_3^{a_3}  D_4^{a_4}  D_5^{a_5}  D_6^{a_6}  D_7^{a_7}  D_8^{a_8}  D_9^{a_9} 
},
\label{eq:FI}%
\end{equation}
where $\TOP \in \{\PL,\NP\}$ labels the families and the denominators
$D_1,\ldots,D_9$ are given in table~\ref{tab:topos}.
We use dimensional regularization with $d=4-2\dimeps$, and our convention for the integration measure for each loop is
\begin{equation}
	\mathfrak{D}^dk = \frac{\td^dk}{\iu\pi^{d/2} \Gamma(1+\dimeps)}.
	\label{eq:loop-measure}%
\end{equation}

\begin{table}[t]
\centering
\begin{tabular}{cl@{\qquad}l}
\toprule
Denominator & integral family $\PL$ & integral family $\NP$ \\
\midrule
$D_1$ & $ k^2 $ &                    $k^2$                   \\
$D_2$ & $l^2 -m_V^2$ &               $(k-l)^2$               \\
$D_3$ & $(k-l)^2$ &                  $(k-p_1)^2$             \\
$D_4$ & $(k-p_1)^2$ &                $(l+p_3)^2-m_V^2$       \\
$D_5$ & $(k-p_1-p_2)^2$ &            $(k-p_1-p_2)^2$         \\
$D_6$ & $(k-p_1-p_2-p_3)^2$ &        $(l-p_1-p_2)^2-m_V^2$   \\
$D_7$ & $(l-p_1-p_2-p_3)^2-m_V^2$ &  $(k-l-p_3)^2$           \\
$D_8$ & $(l-p_1)^2$ &                $(l-p_1)^2-m_V^2$       \\
$D_9$ & $(l-p_1-p_2)^2$ &            $(k-p_1-p_3)^2$         \\
\bottomrule
\end{tabular}
\caption{Definition of the planar ($\PL$) and non-planar ($\NP$) integral families. The loop momenta are denoted by $k$ and $l$, while $m_V$ indicates the mass of the vector boson. The prescription $+ \ieps$ is understood for each propagator and not written explicitly.} \label{tab:topos}%
\end{table}
With the definitions given in Table~\ref{tab:topos}, the two diagrams in figure~\ref{fig:diagrams} can be described using the
first 7 propagators of the two families respectively, and all other diagrams which contribute to the process can be
obtained by permutations of the external gluons and by pinching of the propagators.\footnote{We stress here that if we are interested in computing
the mixed QCD-EW corrections in the $q\bar{q}$ channel, some more integrals are required. We ignore their calculation presently and focus
on the $gg$ channel only.}
Although quite standard, the reduction to a subset of master integrals~\cite{Tkachov:1981wb,Chetyrkin:1981qh,Laporta:2001dd}, 
 is  non-trivial 
 due both to the large number of scales and  the presence of massive internal propagators.
We used \texttt{Reduze2}~\cite{vonManteuffel:2012np} to map the diagrams to the relevant integral families
and performed a complete reduction of all integrals 
with \texttt{KIRA}~\cite{Maierhoefer:2017hyi}.\footnote{
We have also double-checked the IBP-reduction required to derive the differential equations for the master integrals with FIRE5~\cite{Smirnov:2014hma}, see section~\ref{sec:intsDEQs}.}
Finally, we found it  convenient to use FiniteFlow~\cite{Peraro:2019svx} to speed up the substitution of the reduction identities
produced by \texttt{KIRA} in the helicity amplitudes of eq.~\eqref{eq:helampl} and their simplification.

We find that the two independent helicity amplitudes can be expressed in terms of  $116$ master integrals, counting also
the ones obtained through permutations of the external gluons as independent ones.
If we limit ourselves to the un-permuted integrals, we find 43 planar and 18 non-planar master integrals, 
see appendix~\ref{sec:masters} for the full list.  To construct our initial basis of master integrals, we select integrals  whose maximal cuts are defined by integrands with unit leading singularities~\cite{ArkaniHamed:2010gh,Henn:2013pwa,Henn:2020lye}. 
Our choice avoids the appearance of irreducible denominator factors that mix the kinematical variables and the dimensional regularization parameter $d$
during the IBP reduction. This reduces the complexity of intermediate expressions, similarly as described in~\cite{Melnikov:2016qoc}, and recently automated in~\cite{Smirnov:2020quc,Usovitsch:2020jrk}.

In the next two sections we will describe two different strategies that we used to compute the master integrals in terms of multiple polylogarithms.


\section{Computation of the master integrals with differential equations}
\label{sec:intsDEQs}

The standard approach to compute a complete set of multiloop, multiscale Feynman 
integrals goes through deriving and solving their system of differential equations with respect to the masses and momenta, as first worked out in full generality in~\cite{Gehrmann:1999as}.
In each of the invariants $\xi=(s,t,\dots,m^2,\dots)$, a basis of master integrals always fulfils a linear system of differential equations
with rational coefficients. By indicating with $\mathbf{I}$ the vector of master integrals, we can write this system as
\begin{equation*}
	\td\,\mathbf{I}(\xi) =\mathbb{ B }(\dimeps, \xi)\, \mathbf{I}(\xi)\,,
\end{equation*}
where the entries of the matrix $\mathbb{ B }(\dimeps, \xi)$ are differential one-forms that are rational in the kinematics and in the dimensional
regulator $\dimeps$.
One then usually tries to solve these equations as a Laurent series in $\dimeps$, i.e.\ for $d \rightarrow 4$. The effectiveness of this
approach relies on the ability to find a solution of the homogeneous part of the system above, in the limit $\dimeps \to 0$.
While this would be a daunting task given a generic system of coupled differential equations, 
it turns out that an integral representation for the homogeneous solution can always be obtained
by analysing the maximal cuts of the corresponding Feynman integrals~\cite{Primo:2016ebd,Bosma:2017ens,Primo:2017ipr}, 
whose computation becomes particularly simple using the 
so-called Baikov representation~\cite{Baikov:1996iu,Frellesvig:2017aai,Harley:2017qut}.

While the approach described above is completely general, it was shown that in many cases 
the solution of the differential equations can be greatly simplified by the choice of a so-called canonical 
basis of master integrals~\cite{Henn:2013pwa}. If such a basis $\mathbf{F}$ can be found, 
the corresponding system of master integrals
becomes
\begin{equation}
	\td\,\mathbf{F}(\xi) = \dimeps \,\mathbb{A}(\xi)\, \mathbf{F} (\xi)\,, \label{eq:canform}
\end{equation}
where the new matrix $\mathbb{A}(\xi)$ does not depend on $\dimeps$.
In addition to the factorisation of $\dimeps$, an important condition for the basis to be canonical
is that the matrix takes a particularly simple, ``$\td\log$'' form
\begin{equation}
	\label{MBdlog}
	\mathbb{A}(\xi) =  \sum_{j=1}^J A_j\  \td \log P_j\left(\xi\right),
\end{equation}
where $A_j$ are matrices of rational numbers and $P_j$ are algebraic functions of $\xi$, which constitute the \emph{alphabet} $\{P_1,\ldots,P_J\}$ of the problem. 
It follows from eq.~\eqref{MBdlog},
that the master integrals of a canonical $\td\log$ basis can be expressed, order by order in $\dimeps$,
as iterated integrals of the forms $\td\log (P_j$). Furthermore, whenever the alphabet consists entirely of rational functions $P_j$ (or if this can be achieved by an algebraic change of variables), then these iterated integrals can be expressed as linear combinations of the functions
\begin{equation}
G(\sigma_1, \ldots, \sigma_k; x) = \int_0^x \frac{\td \tau}{\tau-\sigma_1} G(\sigma_2,\ldots,\sigma_k; \tau)\,,
\quad
G(\vec{0}_k; x) = \frac{1}{k!} \log^k{x},
\quad
G(;x) = 1\,,
\label{eq:hlog} %
\end{equation}
where the arguments $\sigma_i$ and $x$ will be certain algebraic functions of $\xi$. 
The iterated integrals \eqref{eq:hlog} are known as multiple polylogarithms \cite{Goncharov:2001iea} and hyperlogarithms~\cite{LappoDanilevsky:CorpsRiemann} of \emph{weight} $k$.\footnote{%
	The notation using ``$G$'' was introduced in \cite{Gehrmann:2001jv} as a $G$eneralization of harmonic polylogarithms.}
For most of the Feynman integrals that have been studied so far,
finding a canonical $\td\log$ basis comes along with an expression for the corresponding master integrals
in terms of multiple polylogarithms (with potentially complicated algebraic arguments).
However, no general method to construct an expression of this kind is known if the alphabet cannot be rationalized\footnote{For general algorithms, see for example ~\cite{BesierStratenWeinzierl:Rationalizing}.}
such that, in some cases where a canonical form for the differential equations is known, the
issue of the existence of a polylogarithmic expression for the master integrals remains matter of discussion,
see for example~\cite{Henn:2013woa}.\footnote{One possible approach is the algorithm described in~\cite{Heller:2019gkq}, which is based on an ansatz for the solution.}
In fact, more recently it was shown that there exist iterated integrals
of $\td\log$ forms which \emph{cannot} be expressed in terms of multiple polylogarithms~\cite{Brown:2020rda}. 
In conclusion, whether or not Feynman integrals with a canonical $\td\log$ form can be expressed through multiple polylogarithms, remains an intricate problem.

With these general comments in mind, let us consider now the form of the system of differential equations for the problem at hand.
First of all, it is interesting to notice that, in order to evaluate all the master integrals required for the amplitude, we need to introduce two additional master integrals that would otherwise not appear in our problem, namely $\FIPL{2,2,0,0,0,1,0,0,0}$ and $\FIPL{2,2,0,0,0,1,1,0,0}$. These additional master integrals appear in the non-homogeneous part of the differential equations for some of the top-sector master integrals, and it can immediately be seen that they are obtained by pinching some of the internal lines of the diagrams in Fig.~\ref{fig:diagrams} (see Appendix~\ref{sec:masters} for a complete list of the master integrals).
All master integrals are functions of at most 
four independent variables, which in this section we choose to be $t$, $u$, $m_h^2$, and $m_V^2$ ($V=W,Z$).
Since Feynman integrals are homogeneous functions in the kinematic invariants and in the masses, 
it is convenient to introduce the dimensionless variables
\begin{equation}
	y = -\frac{t}{m_h^2}\,,	\qquad	z = -\frac{u}{m_h^2}\,,	\qquad	\rho = -\frac{m_V^2}{m_h^2}\,,
\end{equation}
in order to factorise the dependence of each master integral on $m_h^2$ as a simple power, namely $(m_h^2)^{d-a_1-\ldots-a_9}$,
where the $a_i$ are the powers of the propagators, see eq.~\eqref{eq:FI}.
For the remainder of this section, we can hence set $m_h^2$ to $1$.
To determine the expressions of the master integrals in terms of the remaining variables we derive differential equations in 
$y$, $z$, and $\rho$ for them and cast this system into a canonical form, as in eq.~\eqref{eq:canform}. This was achieved by starting from
a basis of master integrals whose maximal cuts have unit leading singularities (see appendix~\ref{sec:masters}),
and then applying the algorithm described in~\cite{Gehrmann:2014bfa}.

If we limit ourselves to the 48 planar master integrals, see eq.~\eqref{eq:reduction-basis-PL}, then the differential equations 
take a very simple form and, in particular, all letters are rational functions of $y,z,\rho$ and a single square root,
\begin{equation}\label{eq:rootPL}
R_0 = \sqrt{1+ 4 \rho}
\,.
\end{equation}
As it is well known, this root can be rationalized by the change of variables
\begin{align}
	\label{MBLandau}
	\rho = \frac{1-x}{x^2},
\end{align}
and all integrals of the family $\PL$ can be expressed in terms of multiple polylogarithms whose arguments are rational functions of $x,y,z$, 
as it was shown explicitly in~\cite{Becchetti:2018xsk}, which we
refer to for the explicit form of the differential equations and of the canonical basis.

Unfortunately, even if there is only a small number of new, non-planar integrals, their differential equations 
turn out to be substantially more complicated.
In this case, the vector of planar and non-planar master integrals $\mathbf{F}$ contains 64 entries, and the alphabet
\begin{equation}\label{eq:DE-alphabet}
	\begin{aligned}
	\Big\{& y,
		z,
		 \rho,
		1+y,
		1+z,
		 1+\rho,
		y+z,
		 \rho-y,
		\rho-z,
		1+y+z,
		\rho-y-z,
		\rho+y+z+1,
		\\ &
		\rho-y^2-y,
		\rho-z^2-z,
		\rho-(y+z)^2-y-z,
		R_0,	
		R_1,	
		R_2,	
		R_3,
		R_0+R_1,
		R_0+2y+2z+1,
		\\ &
		 y\,(y+z+1)-\rho z,
		\rho\,(y+1)-z\,(y+z+1),
		\rho\,(y+1)^2-z\,(y+z+1),
		 y z+\rho(y+z)^2,
		 \\ &
		z\,(y+z+1)-\rho y,
		 \rho\,(z+1)-y\,(y+z+1),
		 \rho\,(z+1)^2-y\,(y+z+1),
		 y z-\rho(y+z),
		\\ &
		 R_0+2y+1,
		R_0(y+z)+y-z,
		R_0 (y+1)+y+2z+1,
		 R_1 R_2 y+2\rho\,(1-y+z)-y,
		\\&
		 R_0+2z+1,
		R_1 (y+z)+y-z,
		R_0 (z+1)+2y+z+1,
		 R_1 R_3 z+2\rho\,(1-z+y)-z,
		\\ &  
		R_2+2z+1,	
		R_2+2y+2z+1,
		 y(1+R_2)+2z\,(1+y+z),
		 R_0+R_2,
		1+R_2,
		1+R_1, 
		 \\&
		 R_3+2y+1,	
		R_3+2y+2z+1,
		z(1+R_3)+2y\,(1+y+z),
		 R_0+R_3,
		 1+R_3,	
		 1+R_0
	\Big\}%
	\end{aligned}
\end{equation}
depends on three additional square roots, defined as
\begin{equation}\label{eq:roots-DE}\begin{aligned}
		R_1 &
		     = \sqrt{1 - 4 \rho/(y + z)} 
		\,, \\
		R_2 &
		     = \sqrt{1 + 4 \rho (1 + z) (1 + z/y)}
		\,, \\
		R_3 &
		     = \sqrt{1 + 4 \rho (1 + y) (1+y/z)}
		\,.
\end{aligned}\end{equation}
We provide both the vector of canonical functions $\mathbf{F}$ and the $\td\log$ forms of eq.~\eqref{MBdlog} 
in the ancillary files of this paper. 

As described above, once a canonical form for the differential equations is obtained, the standard procedure consists of constructing 
a solution as a Dyson series in $\dimeps$ whose coefficients consist of iterated integrals.
In case of a three 
scale problem, the usual strategy consists in
solving the partial differential equations sequentially, as described for example in Refs.~\cite{Gehrmann:1999as,Henn:2014lfa}.
We start with one variable and solve the corresponding linear differential equation up to a function of the other two variables, 
then we write down a differential equation with respect to a second variable. We check that the right-hand side of this new equation 
is independent of the first variable and we solve this equation in terms of multiple polylogarithms up to a function of the last variable. 
After the last differential equation is integrated, a solution is obtained up to constants which are then fixed 
by choosing appropriately the boundary conditions. 
While this strategy can be applied straightforwardly when the alphabet is linear
in all variables, 
the presence of several algebraically independent square roots makes it frequently unfeasible
in practice, since at a given step, it is not 
in general possible to find a representation for the result where the corresponding integration variable appears
only in the last argument of the various polylogarithms.

Despite this, it turns out that in our problem 
the four square roots appear in the differential equations
in a very structured pattern, which allows us to devise a solution strategy that is always guaranteed to
terminate and to produce a result in terms of multiple polylogarithms.
First of all, we find it convenient to rationalize the root $R_0$, which appears consistently throughout the 
whole system of equations, by the change of variables of eq.~\eqref{MBLandau}.
We then notice the following crucial structural features of the differential equations:
\begin{itemize}
	\item $R_1$ appears only in the differential equations for the canonical functions $\mathbf{F}_{51}$, $\mathbf{F}_{61}$, $\mathbf{F}_{62}$, $\mathbf{F}_{63}$, $\mathbf{F}_{64}$;
	\item $R_2$ appears only in $\mathbf{F}_{55}$, $\mathbf{F}_{56}$,$\mathbf{F}_{57}$, $\mathbf{F}_{61}$, $\mathbf{F}_{62}$, $\mathbf{F}_{63}$, $\mathbf{F}_{64}$;
	\item $R_3$ appears only in $\mathbf{F}_{58}$, $\mathbf{F}_{59}$,$\mathbf{F}_{60}$, $\mathbf{F}_{61}$, $\mathbf{F}_{62}$, $\mathbf{F}_{63}$, $\mathbf{F}_{64}$;
	\item all the other equations contain at most the root $R_0$;
	\item when solving the equations for $\mathbf{F}_{61}$, $\mathbf{F}_{62}$, $\mathbf{F}_{63}$, $\mathbf{F}_{64}$, at most two square roots are integrated at once and only from weight 3 on: either $\{R_1,R_2\}$ or $\{R_1,R_3\}$.
\end{itemize}
This separation of square roots allows us to perform different changes of variables depending on the canonical functions we want to evaluate, in particular depending on which roots enter a particular integration.

First of all, as customary when dealing with canonical master integrals, we normalise our basis such that all integrals start at order $\dimeps^0$
with a weight $0$ constant (which could of course be zero).
We start by solving the equations for the canonical functions $\mathbf{F}_i$, $i\in\{1,\dots,50,52,53,54\}$, where no square roots appear in $(x,y,z)$. We integrate first in $y$, then in $z$ and at last in $x$, obtaining expressions of uniform weight written in terms of multiple polylogarithms up to some constant factors which will be fixed by imposing boundary conditions. 
We move then to the canonical function $\mathbf{F}_{51}$, where also $R_1$ appears. The relevant letters can be rationalized through the change of variable
\begin{align}
	\label{MBR1rat}
	y = \frac{4(1-x)}{x^2-u^2}-z\,,
\end{align}
and then solved first in $u$ and subsequently in $x$. Only two variables appear here, because $\mathbf{F}_{51}$ is a three-point function.

The triplet $F_{\{55,56,57\}}$ contains the square root $R_2$. We consider the whole subset of functions that enters the differential equation of such triplet, given by $\mathbf{F}_{i}$, $i\in\{1,3,$ $4,6,7,\ldots,12,17,18,\ldots,22,24,25,33,34,35,36,41,46,47,55,56,57\}$, and rationalize $R_2$ by
\begin{align}
	\label{MBR2rat}
	y = \frac{4 (1-x) z (z+1)}{\left(v^2-1\right) x^2-4 (1-x) (z+1)}\,.
\end{align}
The resulting equations are integrated first in $v$, then in $z$, and then in $x$, in terms of multiple polylogarithms.

We proceed in the same way for the triplet $F_{\{58,59,60\}}$, containing $R_3$, and the corresponding subset of canonical functions $\mathbf{F}_{i}$, $i\in\{1,3,4,6,7,\dots,12,17,18,\dots,22,25,26,31,$ $32,35,36,42,44,45,58,59,60\}$. The rationalization of $R_3$ is achieved through
\begin{align}
	\label{MBR3rat}
	z = \frac{4 (1-x) y (y+1)}{\left(w^2-1\right) x^2-4 (1-x) (y+1)}\,.
\end{align}
The order of integration is $w$, $y$, $z$. Also here, the result is expressed in terms of multiple polylogarithms.

The group of canonical functions $\mathbf{F}_{\{61,62,63,64\}}$, corresponding to the master integrals of the top non-planar sector, is the most difficult one. As observed above, up to order $\dimeps^2$ no square roots are present in the variables $(x,y,z)$, therefore the integration in terms of multiple polylogarithms is straightforward, and is carried out following the procedure used for $\mathbf{F}_{\{1,\dots,50,52,53,54\}}$.

Starting from order $\dimeps^3$ all square roots $R_{\{1,2,3\}}$ appear, but always in such a way that a single nested integration contains at most two of them, specifically either $R_1$ and $R_2$, or $R_1$ and $R_3$. We start by removing $R_1$ via the change of variables of eq.~\eqref{MBR1rat}. This change of variables is sufficient to take care of the nested integrations coming from the homogeneous part of the differential equations, as well as of the one coming from the non-homogeneous terms related to $\mathbf{F}_i$, $i\in\{1,\dots,54\}$. Despite the fact that only rational functions are now present for this subset of terms (allowing us to represent this part of the solution in terms of multiple polylogarithms), many cumbersome letters arise, as for example
\begin{equation*}
	z-\frac{4 \left(-u^2 x^3+u^2 x^2+x^5-5 x^4+8 x^3-4 x^2\right)}{u^4 x^2-u^4 x+u^4-2 u^2 x^4+6 u^2 x^3-6 u^2 x^2+x^6-5 x^5+5 x^4}\,.
\end{equation*}
Considering now the terms related to $\mathbf{F}_{\{55,56,57\}}$, a second change of variables to rationalize also $R_2$ is performed ``on the fly'', and reads
\begin{align}
	z = \frac{4 \overline{v} (1-x) x (\overline{v} x-2 x+4)}{\left(x^2-u^2\right) (\overline{v} x+2) (\overline{v} x-2 x+2)+16(1-x)^2}\,.
\end{align}
An analogous ``on the fly'' change of variables is performed on the terms related to $\mathbf{F}_{\{58,59,60\}}$, to get rid of $R_3$:
\begin{align}
	z = \frac{16 \overline{w}^2 (1-x)^2 \left(-u^2+x^2-4 x+4\right)}{\left(u^2-x^2\right) \left(\overline{w}^2 \left(-u^2 x^2+x^4-16 x^2+32 x-16\right)+x^2 \left(u^2-x^2\right)\right)}\,.
\end{align}
Implementing such changes of variables allows us to write the $\dimeps^3$ coefficients of $\mathbf{F}_{\{61,62,63,64\}}$ again in terms of multiple polylogarithms, at the price of having three different sets of independent variables: $(x,u,z)$, $(x,u,\overline{v})$, and $(x,u,\overline{w})$.

To integrate one of the above subsystems, we integrate first in $z$, or $\overline{v}$, or $\overline{w}$, according to preferred variables just discussed.
After the integration in $z$, we verify that plugging these solutions in one of the remaining differential equations gives a matrix of coefficients which is independent of $z$, where this condition must be satisfied considering also the hidden dependence through $\overline{v}$ and $\overline{w}$. The expressions that arise are so cumbersome that we do not see any chance to perform this check analytically. On the other hand, we see numerically (we use {\GiNaC}, \cite{Bauer:2000cp,Vollinga:2004sn} to evaluate multiple polylogarithms) with very high accuracy that our expressions are independent of $z$. However, we cannot simply substitute $z=0$ (which corresponds to $\overline{v}=0$ and $\overline{w}=0$), because individual terms may be singular in this limit. To address this issue, we use shuffle relations to extract carefully all such singular terms as $z \to 0$ explicitly as powers of $\log(z)$. We confirm numerically that the sum of the three contributions in terms of different variables as well as its limit at $z \to 0$ is independent of $z$. Once this is done, we proceed by integrating in $u$, and we check that the remaining differential equation is independent of $u$, in a similar way as we did for $z$.
After having solved all differential equations up to integration constants, we fix these constants using boundary 
conditions in the large-mass limit, $x\to 0$. Here we use well-known prescriptions in a graph-theoretical language for limits typical of Euclidean space -- see, e.g.,~\cite{Smirnov:2002pj}.

The procedure described above can be applied to obtain the $\dimeps^4$ part as well, but the manipulations required are extremely cumbersome 
and, a posteriori, not needed. Indeed, in the next section we will show how to obtain these integrals in a much simpler way by direct
integration over their Feynman/Schwinger parametrisation.
In any case, we believe that the approach we used to solve the differential equations presented here can be used also in other situations where many different square roots appear and only subsets of them are rationalizable at once. The key point of this procedure is to check that in each nested integration only one subset of simultaneously rationalizable square roots is present, and then to perform a ``local'' change of variables ``on the fly'' to rationalize them.


\section{Computation of the master integrals by parametric integration}
\label{sec:finite-basis}

While an expression for the master integrals in terms of multiple polylogarithms can in principle be obtained from the differential equations, the procedure was rather cumbersome as explained in the previous section.
An entirely different approach, which one might attempt, consists in computing all integrals starting from their 
Feynman/Schwinger parametrisation.
This can be in general quite difficult, in particular in multiloop/multileg problems, where one typically needs to integrate over a large number
of Feynman parameters.
Nevertheless, it turns out that in our problem all integrands are linearly 
reducible~\cite{Brown:2008um,Brown:PeriodsFeynmanIntegrals}, which means that the 
algorithms described in \cite{Panzer:2014caa} can be applied rather directly. 

In order to make this approach feasible, it is helpful to choose a basis of master integrals that is finite in the limit $d \to 4$. 
We expect such a change of basis to be  particular useful in the case at hand since the two-loop amplitude is, effectively, a leading-order
amplitude and therefore expected to be finite.
In practice, however, we found it sufficient to replace only the most divergent master integrals with more than $5$ propagators by finite counterparts. To achieve this, we
generated finite integrals by considering the corresponding six-dimensional integrals, including higher powers of the propagators.
For each of the integrals in the families in table~\ref{tab:topos} 
we can  obtain the corresponding $(d+2)$-dimensional integral by~\cite{Tarasov:1996br,Lee:2009dh}
\begin{align}
	\FI^{d+2}_{\TOP}(a_1,a_2,\ldots,a_8,a_9) = \frac{4}{\Delta}
\int \mathfrak{D}^dk\, \mathfrak{D}^dl 
\frac{G(k,l,p_1,p_2,p_3)}{D_1^{a_1} D_2^{a_2}  D_3^{a_3}  D_4^{a_4}  D_5^{a_5}  D_6^{a_6}  D_7^{a_7}  D_8^{a_8}  D_9^{a_9} }
\label{eq:dp2int}%
\end{align}
where $G(p_1,\ldots,p_n)$ is the Gram-determinant of the $n$ momenta $p_1,\ldots,p_n$, and
\begin{equation*}
\Delta =  G(p_1,p_2,p_3) (d-4)(d-3) = \frac{s\,t\,u}{4} (d-4)(d-3)\,.
\end{equation*}
It is pretty easy to see that, at least in the case at hand, 
as long as we choose UV finite integrals and all powers of the massless propagators equal to unity, 
the Gram determinant  $G(k,l,p_1,p_2,p_3)$ cures all IR divergences,
both in the collinear and in the soft limits. This allows us to easily generate  a large number of finite integrals.
We stress that this is particularly straightforward here due to the presence of two internal massive propagators. In fact, 
even for integrals with fewer propagators (and therefore with poor UV behaviour), we can simply raise the powers of the massive
propagators \emph{ad libitum} in order to
obtain UV-finite integrals, without spoiling their IR behaviour. We note that, in a general case,  finite
integrals can be found algorithmically also in the absence of massive propagators, see for example the algorithm 
described in~\cite{vonManteuffel:2014qoa}.
We list the finite integrals used in this calculation in Appendix~\ref{sec:finitemasters}.

\subsection{Planar integrals}
\label{sec:intsHyperInt}

The parametric representation \cite{Nakanishi:GraphTheoryFeynmanIntegrals,Smirnov:AnalyticToolsForFeynmanIntegrals} 
of an integral family such as \eqref{eq:FI} has the form
\begin{equation}
	\FI_{\TOP}^d(a_1,\ldots,a_9)
	=
	\frac{(-1)^{\sdc+d}\Gamma(\sdc)}{\Gamma(1+\dimeps)^2}
	\left(\prod_{k=1}^9 \int_0^{\infty} \frac{x_k^{a_k-1} \td x_k}{\Gamma(a_k)} \right)
	\frac{\delta(1-x_j)}{\U^{d/2-\sdc}_{\TOP} \F^{\,\sdc\phantom{/}}_{\TOP}},
	\label{eq:FI-parametric}%
\end{equation}
where $\sdc = a_1+\cdots+a_9-d$. The polynomials
$\U = \det A$ and $\F = \U (B^{\Transpose} A^{-1} B - C)$
are determined by the quadratic ($A$), linear ($B$) and constant ($C$) parts of the quadratic form
\begin{equation*}
	x_1 D_1+\cdots + x_9 D_9
	=
	\ell^{\Transpose}\!A\,\ell
	+2 B^{\Transpose} \ell
	+C
\end{equation*}
in the two loop momenta $\ell=\left(\begin{smallmatrix} k \\ l \\\end{smallmatrix}\right)$, given by the denominators from Table~\ref{tab:topos}.
All integrals that we are interested in for this calculation, for both integral families, are chosen such that $a_8=a_9=0$, which allows us to
eliminate the parameters $x_8$ and $x_9$. 
The remaining denominators $D_1,\ldots,D_7$ are the inverse scalar propagators of the graphs shown in Figure~\ref{fig:diagrams}. 
Concretely, in the planar case we find the Symanzik polynomials to be
\begin{equation}\label{eq:UF-PL}\begin{split}
	\U &= x_3 (x_1+x_2+x_4+x_5+x_6+x_7) + (x_2+x_7)(x_1+x_4+x_5+x_6) \quad\text{and}\\
	\F &= -m_h^2 \big(
		(x_1x_2+x_1x_3+x_2x_3)(x_6+x_7)+x_6 x_7 (x_1+x_2)+x_2 x_7(x_4+x_5)
	\big) \\
	&\quad -s x_5\left( x_1x_2 + x_1x_3 + x_1x_7 + x_2x_3 \right)
	       -u x_4\left( x_2x_6 + x_3x_6 + x_3x_7 + x_6x_7 \right) \\
	&\quad  + m_V^2 (x_2+x_7)\,\U.
\end{split}\end{equation}
An analysis by polynomial reduction \cite{Brown:PeriodsFeynmanIntegrals} shows that the set $\{\U,\F\}$ is \emph{linearly reducible}. This means that the integrals \eqref{eq:FI-parametric} 
can be expressed algorithmically in terms of the hyperlogarithms defined in eq.~\eqref{eq:hlog}. In fact, this works to all orders of the $\dimeps$ expansion, and for arbitrary integer values of $a_1,\ldots,a_7$.

The algorithm described in~\cite{Brown:2008um} applies directly only to convergent integrals. 
As explained above, we therefore adjusted our basis to consist mostly of finite integrals. The remaining divergences in this basis (see Appendix~\ref{sec:finitemasters}) occur only in integrals with $4$ or fewer propagators, and six further integrals with $5$ or $6$ propagators, where they can be resolved easily
through integration by parts in the parameters $x_k$, following the method of \cite{vonManteuffel:2014qoa,Panzer:DivergencesManyScales}.
In order to perform the polynomial reduction, resolution of divergences, and integration over the Feynman parameters explicitly, we used the code {\HyperInt} \cite{Panzer:2014caa}. 
Starting from the polynomials \eqref{eq:UF-PL}, {\HyperInt} identifies $x_1,x_4,x_5,x_6,x_3$ 
as an admissible order for the first five integrations. They result in expressions with hyperlogarithms 
whose arguments $\sigma_i$ are rational functions of $s,t,u,m_V^2$ and $x_2,x_7$. 
We pick $j=7$ for the constraint $x_7=1$ in \eqref{eq:FI}, leaving the final integral over $x_2$. 
At this stage, the algorithm needs to solve for the roots of the polynomial $m_V^2(1+x_2)^2 - m_h^2 x_2$, which introduces
the first square root
\begin{equation}\label{eq:root-PL}
	r 
	= \sqrt{m_h^2(m_h^2-4m_V^2)}
	= (-m_h^2) R_0,
\end{equation}
which we saw also in \eqref{eq:rootPL}.
Consequently, the final expressions for the coefficients of the $\dimeps$-expansion of the 
integrals $\FI_{\PL}$ are  linear combinations of hyperlogarithms, whose coefficients and arguments 
are rational functions of $s,t,u,m_V^2$ and $r$.

\subsection{Non-planar integrals}
For the non-planar integral family, the corresponding polynomials are
\begin{equation}\label{eq:UF-NP}\begin{aligned}
	\U &= (x_1+x_3+x_5)(x_2+x_4+x_6+x_7) + (x_2+x_7)(x_4+x_6),\\
	\F &= m_V^2 (x_4+x_6)\U  -m_h^2 \big(x_1x_6(x_4+x_7)+x_2x_4x_5 +x_4x_6(x_2+x_3+x_5+x_7)\big)\\
	   &\quad -s \big( x_1x_5(x_2+x_4+x_6+x_7)+x_1x_2x_6+x_4x_5x_7 \big)
	          -t x_2x_3x_4 
		  -u x_3x_6x_7,
\end{aligned}
\end{equation}
and it was pointed out in \cite[Figure~10]{Panzer:DivergencesManyScales} that they are linearly reducible too. As an admissible order for the first integrations we use $x_1,x_3,x_5,x_2,x_7$. Setting $x_6=1$, the final integration over $x_2$ introduces three further square roots in addition to $r$:
\begin{equation}\label{eq:roots-NP}
	\sqrt{1-4m_V^2/(t+u)}=R_1,\ 
	\sqrt{ r^2-4m_V^2 su/t}=-m_h^2 R_2,\ 
	\sqrt{ r^2-4m_V^2 st/u}=-m_h^2 R_3,
\end{equation}
which we encountered also in the differential equations, see \eqref{eq:roots-DE}.
Our results for the integrals $\FI_{\NP}$ from the basis \eqref{eq:finitebasis} therefore 
consist of linear combinations of hyperlogarithms with coefficients and arguments that are
 rational functions of $s,t,u,m_V^2,r$ and the three roots in \eqref{eq:roots-NP}.
 In fact, the polynomial reduction shows that the quadratic polynomials responsible for $R_2$ and $R_3$ are not \emph{compatible} \cite{Brown:PeriodsFeynmanIntegrals} with each other.
Explicitly, this manifests itself in the fact that our results admit a decomposition
\begin{equation}
	A\left(s,t,u,m_V^2,r,R_1,R_2\right)
	+ B\left(s,t,u,m_V^2,r,R_1,R_3\right)
	\label{eq:root-separation}%
\end{equation}
into expressions $A$ and $B$ whose hyperlogarithm arguments $\sigma_k$ in \eqref{eq:hlog} are rational functions of the listed arguments only, i.e.\  the roots $R_2$ and $R_3$ do not mix.
This property corresponds to the structure of the differential equations
described in section~\ref{sec:intsDEQs}, and makes it possible to rationalize the pieces $A$ and $B$ individually. 
For the parametric integration, however, such rationalizations provide no advantage. 
In contrast, the bare expressions with the (unrationalized) roots are much more compact.
\begin{remark}
	We stress that the hyperlogarithm expressions obtained from {\HyperInt} are valid for all values of the kinematic parameters such that the integral \eqref{eq:FI-parametric} converges. In particular, by giving a small positive imaginary part to $s,t$ and $u$ in order to implement the $\ieps$ prescription, these hyperlogarithms can be evaluated directly in the physical region, for example using {\GiNaC} \cite{Vollinga:2004sn}. 
	This is a very valuable property, because the analytic continuation of polylogarithms with
	algebraic arguments is typically much more delicate.
\end{remark}

For all $\dimeps$-expansion coefficients of the integrals \eqref{eq:finitebasis} that 
contribute to the helicity amplitudes \eqref{eq:helampl}, we find that only hyperlogarithms of weight $k\leq 4$ arise. 
This weight bound is consistent with other known two-loop amplitudes in four dimensions.
In ancillary files to this publication, we provide the explicit 
expressions thus obtained for all coefficients of the $\dimeps$-expansions of the 
integrals in our basis \eqref{eq:finitebasis} that are required for the 
computation of the helicity amplitudes. 
The ancillary files also include instructions and code to 
reproduce these calculations.


\section{The helicity amplitudes}
\label{sec:results}

Combining our results for the Feynman integrals, we obtain expressions for the helicity amplitudes $\Omega_{++\pm}^{(0)}$. 
At this step, we see that all poles in $\dimeps$ stemming from individual divergent integrals, 
as well as from the coefficients in the reduction of the amplitudes to the Feynman integrals, 
completely cancel each other. As expected,
the helicity amplitudes thus turn out to be finite.
Furthermore, we notice that:
\begin{itemize}
	\item In the case of $\Omega_{+++}^{(0)}$, all hyperlogarithms of weight $4$ cancel out, leaving only functions of weight at most 3 in the result. 
A similar weight drop was found in mixed QCD-EW corrections to $gg \to H$, see~\cite{Bonetti:2016brm,Bonetti:2017ovy}, 
where the two- and three-loop amplitudes turn out to have maximum weight three and five, respectively.
	\item In the case of $\Omega_{++-}^{(0)}$, hyperlogarithms of weight $4$ do not cancel completely and persist in the result.
\end{itemize}
These weights may at first seem surprising, in particular because no such weight drop shows up 
in the corresponding Higgs 
Effective Field Theory (HEFT) amplitudes, see for example~\cite{Gehrmann:2011aa}.
But for our mixed QCD-EW corrections the weight drop can be a posteriori justified, rather heuristically, as follows.
If we consider the possible unitarity cuts of the $\Omega_{+++}^{(0)}$ helicity amplitude, 
we find that supersymmetric ward identities ensure that
all cuts which go through two massless quark lines
are zero in $\dimeps=0$, while non-zero contributions are only obtained cutting through at 
least one of the massive vector bosons.
We expect that the weight four part of the amplitude should be proportional 
to the master integrals whose coefficients can be obtained by projecting over the former type of cuts. 
Their vanishing in
$d=4$ can therefore be seen as an argument in favour of the observed weight drop. 
Clearly, the same argument
applies equally well to $gg \to H$, where the only helicity amplitudes different from 
zero are for equal-helicity gluons.
On the other hand, this reasoning fails $\Omega_{++-}^{(0)}$ and no weight drop is observed.

After some simplification, our result for the helicity amplitude $\Omega_{+++}^{(0)}$ takes the form
\begin{equation}
	\Omega_{+++}^{(0)}(s,t,u) 
	= -16 
	+ \frac{4 m_V^2}{m_h^2} \Big[ H(s,t,u)+H(t,s,u)+H(u,t,s) \Big]\,,
\end{equation}
where the hyperlogarithms $H=H_1+H_2+H_3$ are given in weight $1$ and $2$ explicitly as
\begin{align}
	H_1(s,t,u)
	&=
	\left(1-\frac{s}{m_V^2}\right)\log\left(1-\frac{s}{m_V^2}\right)
	-\left(1-\frac{m_h^2}{m_V^2}\right) \log\left(1-\frac{m_h^2}{m_V^2}\right)
	+\frac{2r}{3m_V^2} \log \frac{r-m_h^2}{r+m_h^2}
	\nonumber\\
	H_2(s,t,u)&=
	-\left(2+\frac{su}{m_V^2 t}+\frac{st}{m_V^2 u}\right)
	\Li{2}\left( \frac{t+u}{m_h^2-m_V^2} \right)
	-\left(1+\frac{su}{m_V^2 t}+\frac{st}{m_V^2 u}\right)
	\Li{2}\left(\frac{s}{m_V^2}\right)\nonumber\\&\quad
	+\left( 1+\frac{tu}{m_V^2 s} \right)
	\bigg(
		\Li{2}\left( \frac{s+ut/m_V^2}{m_h^2-m_V^2} \right)
		-\frac{1}{2}\log^2\left(1-\frac{m_h^2}{m_V^2}\right)
	\bigg).
	\label{eq:ppp-weights12}%
\end{align}
The expression for $H_3$, the hyperlogarithms of weight 3, is provided in the ancillary files.
Their arguments are rational functions of $s,t,u,m_V^2$ and the roots
\begin{equation}\label{eq:roots+++}\begin{aligned}
	r   &= (-m_h^2)\sqrt{1-4m_V^2/m_h^2}, &
	r_s &= \sqrt{ r^2-4m_V^2 ut/s}, \\
	r_t &= \sqrt{ r^2-4m_V^2 su/t}\qquad\text{and}\quad &
	r_u &= \sqrt{ r^2-4m_V^2 st/u}.
\end{aligned}\end{equation}

The amplitude $\Omega_{++-}^{(0)}$ is more complicated, not only because it involves hyperlogarithms of weight 4, but
also since their arguments  require two additional square-roots,
\begin{equation}\label{eq:roots++-}
	r_{st} = \sqrt{1-4m_V^2/(s+t)} \qquad\text{and}\qquad
	r_{su} = \sqrt{1-4m_V^2/(s+u)}.
\end{equation}
These roots arise from $R_1$ in the crossed versions ($s\leftrightarrow t$ or $s\leftrightarrow u$) of the integrals that we computed in section~\ref{sec:intsDEQs} and section~\ref{sec:finite-basis}. Similarly, crossing is responsible for the appearance of the root $r_s$ in \eqref{eq:roots+++}. The explicit form of $\Omega_{++-}^{(0)}$ is provided in the ancillary files.

The results so obtained can be evaluated rather straightforwardly in any region of phase-space, in particular 
both in the Euclidean, $s,t,u<0$, and in the physical\footnote{%
	The physical values of the boson masses fix $m_h^2/m_V^2$ to either $\approx 2.425$ ($V=W$) or $\approx 1.885$ ($V=Z$), so in particular, $m_V^2<m_h^2<4m_V^2$ is fulfilled.} Minkowskian region where
\begin{equation}
	\label{eq:physical-region}%
	t,u<0<m_V^2< m_h^2 < 4 m_V^2
	\quad\text{and hence}\quad s > m_h^2=s+t+u.
\end{equation}
Indeed, the hyperlogarithms can be evaluated numerically with {\GiNaC} \cite{Vollinga:2004sn}, provided a small imaginary part is given to $s$, $t$ and $u$.
This is needed, also in the Euclidean region, because individual hyperlogarithms in the expression are not necessarily single-valued, and a
consistent determination for all of them must be picked. 
In the Euclidean region, all choices for the signs of the infinitesimal imaginary parts produce the same, real, result.
The correct result in the physical region, however,
is obtained by ensuring that both $s$ and $m_h^2 = s+t+u$ have a positive imaginary part (according to the $\ieps$ prescription).

For reference, table~\ref{tab:numval} provides numerical results for the helicity amplitudes in two points in the Euclidean region and two points (plus two crossings) in the physical region. We pick two points in the \emph{physical region} \eqref{eq:physical-region}
with $m_h^2 = (125/90)^2 m_V^2$ such that
\begin{equation}\label{eq:samples-phys}
p_1^{\textup{ph}} =\left\{ s \to \frac{1225}{324}m_V^2\,,\; t \to - \frac{25}{81}m_V^2  
\right\} \,, \quad 
p_2^{\textup{ph}} =\left\{ s \to \frac{937}{324}m_V^2\,,\; t \to - \frac{275}{324}m_V^2 
\right\}\,,
\end{equation}
and two points in the \emph{Euclidean region} with
\begin{equation}\label{eq:samples-euc}\begin{aligned}
p_1^{\textup{eu}} &= \left\{ s \to -\frac{1225}{324}m_V^2\,,\; t \to - \frac{25}{81}m_V^2 \,,\; u \to - \frac{500}{324}m_V^2 
\right\} \,, \\
p_2^{\textup{eu}} &=\left\{ s \to -\frac{937}{324}m_V^2\,,\; t \to - \frac{275}{324}m_V^2 \,,\; u \to - \frac{37}{324}m_V^2 
\right\}\,.
\end{aligned}\end{equation}

\begin{table}\centering
    \begin{tabular}{ccc}
      \textbf{point in phase-space} & \boldmath $ \Omega_{+++}^{(0)}\left(s,t,u,m_V^2\right)$ & \boldmath $\Omega_{++-}^{(0)}\left(s,t,u,m_V^2\right)$ \\[0.1cm]
      \toprule
      $p_1^{\textup{ph}}$ & ~$-7.2015542 - 0.8783012 \,\iu $~ & ~$ -6.6933149 - 
 0.9982990 \,\iu$~ \\[1mm]
	$p_1^{\textup{ph}}\text{\rlap{\ $(s \leftrightarrow t)$}} $ & ~$-7.2015542 - 0.8783012 \,\iu $~ & ~$-14.7100602 -13.1607693 \,\iu$~ \\[1mm]
	$p_1^{\textup{ph}}\text{\rlap{\ $(s \leftrightarrow u)$}} $ & ~$-7.2015542 - 0.8783012 \,\iu $~ & ~$ -6.8240815 - 1.5445788 \,\iu$~ \\[1mm]
      $p_2^{\textup{ph}}$ & ~$-7.1894251 - 0.7143046 \,\iu $~ & ~$-7.0299953 - 
 0.8180534 \,\iu$~ \\[0.08cm]
      \midrule
      $p_1^{\textup{eu}}$ &  ~$ 5.3960378$~ &   ~$5.3720766$~   \\[0.1cm]
      $p_2^{\textup{eu}}$ &  ~$4.3872778 $~ &   ~$4.3814106$~ 	\\[0.08cm]
      \bottomrule
    \end{tabular}
    \caption{Numerical values for the two helicity amplitudes in the Euclidean and in the physical region, at the points defined in eqs.~\eqref{eq:samples-phys} and \eqref{eq:samples-euc}.}%
        \label{tab:numval}%
\end{table}

It is interesting to notice that, in a rather large portion of the physical phase-space (i.e. $s > m_h^2$), 
the two helicity amplitudes are 
numerically similar. This is in part due to the fact that the two amplitudes are expected to go
to the same value both in the limit $m_V \to \infty$ and when  the gluon $p_3$ becomes soft, see section~\ref{sec:checks} for details. 

For completeness, in table~\ref{tab:numval} we also show the value of the helicity amplitudes in the 
physical region but for crossed kinematics.  Clearly 
$\Omega_{+++}$ is symmetric under $p_i \leftrightarrow p_j$, while $\Omega_{++-}$ is not and, instead, the two crossings $p_1 \leftrightarrow p_3$ and $p_2 \leftrightarrow p_3$
correspond to the missing helicity amplitudes $\Omega_{-++}$ and $\Omega_{+-+}$, respectively.

\subsection{Polylogarithm expressions for $\Omega_{+++}^{(0)}$}
While the amplitudes in the form discussed above are guaranteed to produce the correct result, if 
the Feynman $\ieps$ prescription is applied, the numeric evaluation of the hyperlogarithms 
is not particularly efficient, especially in the physical region.
In order to obtain a fast and stable method to evaluate the helicity amplitudes, we rewrite
the hyperlogarithms in terms of simpler functions. In particular, classical polylogarithms \cite{Lewin:PolylogarithmsAssociatedFunctions} of weight $k$,
\begin{equation}
	\Li{k}(z) = \sum_{n=1}^{\infty} \frac{z^n}{n^k}
	\qquad\text{for}\qquad \abs{z}<1,
	\label{eq:Li}%
\end{equation}
are readily available for speedy evaluation in many computer algebra systems. It was demonstrated 
in \cite{Kummer:IntegrationenRationalerFormeln} that every hyperlogarithm of weight $3$ can be 
expressed as a linear combination of $\Li{3}$'s with suitable arguments, plus products of $\Li{2}$'s 
and logarithms. However, in deriving such an expression for $\Omega_{+++}^{(0)}$, great care is required due to the multi-valuedness of polylogarithms.
The principal branches have discontinuities on the rays
\begin{equation}\label{eq:branch-cuts}
(-\infty,0]\quad\text{for}\quad \log, \quad\text{and}\quad
[1,\infty) \quad\text{for}\quad\Li{k}.
\end{equation}
An expression built out of principal branches of polylogarithms typically develops discontinuities whenever an argument crosses one of these branch-cuts. It is therefore not always possible to find a single expression that captures the desired branches over the entire  phase-space. Instead, different expressions must be derived 
in various sub-regions of phase-space. In the ancillary files, we therefore provide two different expressions for $\Omega_{+++}^{(0)}$ written in terms of $\Li{3},\Li{2}$ and logarithms only:
\begin{itemize}
	\item one expression is valid in the entire Euclidean region defined by $s,t,u<0<m_V^2$.
	\item one expression is valid in the entire physical region defined in eq.~\eqref{eq:physical-region}.
\end{itemize}
Note that due to the symmetry of $\Omega_{+++}^{(0)}$ under permutations of $s,t$ and $u$, 
the latter region completely determines this helicity amplitude in the entire physical region of interest.

In the Euclidean case, the four roots \eqref{eq:roots+++} are positive and real. The arguments of the polylogarithms $\Li{2}$ and $\Li{3}$ in our expression are chosen to be real and less than $1$, over the entire Euclidean region.
Hence, the resulting expression is manifestly real in the entire Euclidean region and efficient to evaluate.

After analytic continuation, in the physical region the roots \eqref{eq:roots+++} take the values
\begin{equation}
	r = -\iu \cdot \abs{r},\quad
	r_s = -\iu \cdot \abs{r_s},\quad
	r_t = -\iu \cdot \abs{r_t},\quad
	r_u = -\iu \cdot \abs{r_u},
	\label{eq:roots+++phys}%
\end{equation}
and we ensured that the arguments of all polylogarithms in our corresponding expression stay away 
from the branch cuts \eqref{eq:branch-cuts}, throughout the entire region \eqref{eq:physical-region}.
Our second polylogarithm expression for $\Omega_{+++}^{(0)}$, tailored for the physical region and given in the ancillary files, can thus be evaluated in that region efficiently and robustly, without any ambiguities.

\begin{remark}
	A priori, it is not guaranteed that such an expression, single-valued throughout the entire physical region, even exists at all. Further subdivisions of phase-space might have been required, see for example \cite{Gehrmann:2014bfa,Gehrmann:2015ora,Heller:2019gkq}.
\end{remark}

In order to derive the expressions for $\Omega_{+++}^{(0)}$ discussed above, we followed roughly the approach outlined in \cite{Duhr:2011zq}. First, we computed the symbol of the amplitude, which we find to produce 39 letters, namely
\begin{equation}\label{eq:+++alphabet}\begin{gathered}
	\frac{s}{m_V^2}, 1-\frac{s}{m_V^2},
	1+\frac{m_V^2s}{tu},
	1-\frac{s(t+u)}{m_h^2 m_V^2},
	1-\frac{r}{m_h^2}, 1-\frac{r+2u}{m_h^2},
	\frac{r+r_s}{m_h^2},\phantom{-\frac{2m_V^2}{m_h^2}}
	\\
	\frac{m_h^2}{m_V^2}, 1-\frac{m_h^2}{m_V^2},
	\frac{s+t}{m_V^2},
	1-\frac{m_V^2(t+u)}{tu}, 
	1-\frac{r_s}{m_h^2},
	 1-\frac{r_s+2u}{m_h^2},
	 1-\frac{r_s+2m_V^2}{m_h^2},
\end{gathered}\end{equation}
and their conjugates under crossings. We then transformed this symbol into a basis of Lyndon words \cite{MelanconReutenauer:LyndonFree}, which separates the $\Li{3}$-contributions from the products of $\Li{2}$'s and $\log$'s. Finally, we matched this expression to an ansatz of $\Li{3}(z)$'s, where the arguments $z$ are constructed such that:
\begin{itemize}
	\item $z$ and $1-z$ both factorize over the alphabet (given by \eqref{eq:+++alphabet} and conjugates),
	\item $z$ never crosses the branch-cut $[1,\infty)$.
\end{itemize}
The second condition selects different arguments for the Euclidean and physical regions, leading to different final expressions. To check for the factorizations in the first condition, we used integer relation techniques as detailed in \cite[section~3]{ManteuffelTancredi:NP2loop3pt}.

For the other helicity amplitude $\Omega_{++-}^{(0)}$, the result includes hyperlogarithms of weight up to and including four,
and the corresponding symbol alphabet is more involved due to the presence of the two extra square roots in~\eqref{eq:roots++-}.
In a similar way as above, it would be possible to rewrite our expressions in terms of simpler polylogarithms, reducing the set of transcendental functions 
to $\log,\Li{2},\Li{3},\Li{4}$ and $\Li{2,2}$, as explained for example in \cite{FrellesvigTommasiniWever:Li22}. 
We leave this to future work.


\subsection{Checks on the result}
\label{sec:checks}

Each master integral, with the exception of the weight four piece of the 7-propagator non-planar integrals, 
has been successfully checked using the \texttt{Mathematica} \cite{MathematicaProg} package \texttt{PolyLogTools} \cite{Duhr:2019tlz,Bauer:2000cp,Vollinga:2004sn} to numerically compare its expression obtained via differential equations to its expression calculated through integration over Feynman parameters in multiple points inside the Euclidean region. Furthermore, the results from Feynman parameters integration (including weight four for the 7-propagator integrals) have been checked numerically against \texttt{PySecDec} \cite{Borowka:2017idc,Borowka:2018goh,Hahn:2004fe,Hahn:2014fua,Kuipers:2013pba,Ruijl:2017dtg,0954161734} both in the Euclidean and in the Minkowski region, finding excellent agreement in all points. Finally, also
the results from the differential equations have been checked in random points in the 
Euclidean region against \texttt{FIESTA} \cite{Smirnov:2015mct}, finding excellent numerical agreement.

~

To validate our results for the amplitude we considered two different limits for the amplitude: the soft-gluon limit and the limit of a vector boson with infinite mass.

In the soft limit, the $gg \to H g$ amplitude $\mathcal{A}_{\lambda_1 \lambda_2 \lambda_3}^{c_1 c_2 c_3}$ factorizes into the leading order $gg \to H$ amplitude $\mathcal{A}_{\lambda_1 \lambda_2}$ times an eikonal factor.\footnote{The color structure of the leading order amplitude has been included in the eikonal factor.} Using the gauge choice of eq.~\eqref{eq:gauge}, the factorization takes the form
\begin{align}
	\mathcal{A}_{\lambda_1 \lambda_2 \lambda_3}^{c_1 c_2 c_3} \xrightarrow[p_3 \to 0]{} -\iu g_s f^{c_1 c_2 c_3} \frac{p_2 \cdot \epsilon_{\lambda_3}}{p_2 \cdot p_3} \mathcal{A}_{\lambda_1 \lambda_2}\,,
\end{align}
which can be rewritten in terms of spinor products as
\begin{align}
	\begin{aligned}
		\mathcal{A}_{+++}^{c_1 c_2 c_3} &\xrightarrow[p_3 \to 0]{} -\iu g_s f^{c_1 c_2 c_3} \sqrt{2} \frac{\langle12\rangle}{\langle13\rangle\langle23\rangle} \mathcal{A}_{++}\,,	\\
		\mathcal{A}_{++-}^{c_1 c_2 c_3} &\xrightarrow[p_3 \to 0]{} -\iu g_s f^{c_1 c_2 c_3} \sqrt{2} \frac{[12]}{[31][32]} \mathcal{A}_{++}\,.
	\end{aligned}
\end{align}
Using the same normalisation for the EW and QCD couplings, the 
leading order amplitude for $gg \to H$ for gluons of plus helicity can be written schematically as \cite{Bonetti:2017ovy}
\begin{align}
	\mathcal{A}_{\lambda_1 \lambda_2} = \epsilon_{\lambda_1} \cdot \epsilon_{\lambda_2} \,
	\mathcal{F}\left( \frac{m_h^2}{m_V^2}\right) \,,\quad \mbox{such that} \quad \mathcal{A}_{++}= \frac{[12]}{\langle12\rangle} \mathcal{F}\,,
\end{align}
where $\mathcal{F}$ is a non-trivial function of the ratio $m_h^2/m_V^2$.
Inserting the expression above into the soft limit we get
\begin{align}
	\begin{aligned}
		\mathcal{A}_{\textup{soft},+++}^{c_1 c_2 c_3} &= 
		-\iu g_s f^{c_1 c_2 c_3} \sqrt{2} \frac{m_h^2}{\langle12\rangle\langle23\rangle\langle31\rangle} \, \mathcal{F}\,,	\\
		\mathcal{A}_{\textup{soft},++-}^{c_1 c_2 c_3} &= 
		-\iu g_s f^{c_1 c_2 c_3} \sqrt{2}\frac{[12]^3}{[13][23]m_h^2}\, \mathcal{F}\,,
	\end{aligned}
\end{align}
which correspond to our expressions for the amplitude in eq.~\eqref{eq:helamplit}. 
Indeed, we could check numerically that for $t \to 0^-$, $u \to 0^-$, $s \to m_h^2$, we obtain
\begin{align}
\lim_{p_3 \to 0} \Omega_{++-}^{(0)} \; = \;  \lim_{p_3 \to 0}  \Omega_{+++}^{(0)}\; = \; \mathcal{F}\,.
\end{align}

~

To check the $m_V \gg m_h $ limit we start by recalling that, in this approximation, the interaction can
be encapsulated in a Wilson coefficient for the effective Lagrangian \cite{Anastasiou:2008tj,Gehrmann:2011aa}
\begin{equation}
	\mathcal{L}_{\textup{eff}} = - \alpha_s \frac{C_1}{4 v} H G_{\mu\nu}^a G^{\mu\nu}_a\,,
\end{equation} 
where $v$ denotes the vacuum expectation value of the Higgs field.\footnote{We should note here that in the case of $gg \to Hg$
also another operator could appear which couples the Higgs boson directly to a $q\bar{q}$ pair and a gluon. 
We do not consider this operator here, since it is suppressed by one more power in $1/m_V^2$ in the limit $m_V \to \infty$.}
Up to the explicit form of the
Wilson coefficient $C_1$, this Lagrangian is identical to the heavy-top mass Lagrangian. We can therefore
read off the leading order mixed QCD-EW $gg \to Hg$ amplitude directly from the corresponding 
computation in the heavy-top limit, which is presented in~\cite{Gehrmann:2011aa} as
\begin{align}
	\begin{aligned}
		\mathcal{A}_{\textup{eff},+++}^{c_1 c_2 c_3} &= \alpha_s \frac{C_{1,\textup{EW}}}{v} \sqrt{4\pi\alpha_s}f^{c_1 c_2 c_3} \frac{m_h^4}{\sqrt{2}\langle12\rangle\langle23\rangle\langle31\rangle},	\\
		\mathcal{A}_{\textup{eff},++-}^{c_1 c_2 c_3} &= \alpha_s \frac{C_{1,\textup{EW}}}{v} \sqrt{4\pi\alpha_s}f^{c_1 c_2 c_3} \frac{[12]^3}{\sqrt{2}[23][13]},
	\end{aligned}
\end{align}
with $C_{1,\textup{EW}} = - \alpha \left(C_W + C_Z \cos^2\theta_W\right) / \left(16\pi^2 \sin^2\theta_W\right)$.
In order to compare these to our results, we expand our helicity amplitudes in the limit $m_V\rightarrow \infty$. We find that both helicity amplitudes agree at leading order in this limit, such that\footnote{We verified these expansions symbolically for $\Omega_{+++}^{(0)}$ and numerically for $\Omega_{++-}^{(0)}$. They are valid in all regions of phase-space.}
\begin{equation}\label{eq:omega-mV=infinity}
	\Omega^{(0)}_{++\pm}(s,t,u,m_V^2) = -2 \frac{m_h^2}{m_V^2} + \mathcal{O}\left(\frac{1}{m_V^4}\right)
	.
\end{equation}
We then infer the corresponding expressions for $\mathcal{A}_{\textup{eff},+++}^{c_1 c_2 c_3}$ and $\mathcal{A}_{\textup{eff},++-}^{c_1 c_2 c_3}$ via \eqref{eq:helamplit}, and find agreement with \eqref{eq:omega-mV=infinity} after multiplying our results by a factor of $\iu$.

For future applications, we note that the soft limit $p_3\rightarrow 0$ showed that there is \emph{no} relative phase factor between our results for $gg\rightarrow Hg$, and $gg\rightarrow H$ as given in \cite{Bonetti:2017ovy}.


\section{Conclusions}
\label{sec:conc}

In this paper we described the first calculation of the two-loop mixed QCD-EW corrections to 
the production of a Higgs
boson and a gluon in gluon fusion through a loop of massless quarks, 
with full dependence on the Higgs and on the vector boson masses. 
The amplitudes presented here are the last missing building blocks 
required to compute the NLO mixed QCD-EW corrections to Higgs production in gluon fusion,
overcoming the shortcoming of the various approximations that have been used to estimate these
corrections in the past.
We made use of helicity projector operators to extract the two independent
helicity amplitudes from the two-loop Feynman diagrams that contribute to the process in terms
of scalar Feynman integrals.  We  reduced all scalar integrals to master integrals by  use of integration
by parts identities and computed the master integrals with two independent methods, namely 
both starting from their differential equations in canonical form and by direct integration over their
Feynman/Schwinger parametrisation. In both cases, we find that the result can be expressed in terms
of multiple polylogarithms. 
Achieving this form by integrating the differential equations turned out to be 
cumbersome in practice, in spite of the fact that a canonical form for the differential equations could be found. 
In fact,  the alphabet of the non-planar master integrals
is characterised by the presence of four independent square roots,
that we did not manage to rationalize at the same time. For this reason, integrating the equations 
required us to split the master integrals
into different contributions, and to use different changes of variables to rationalize the square roots
in each of these pieces. This was doable in practice thanks to the particular
structure of the system of differential equations, but it produced rather cumbersome results.

Interestingly, the fact that all integrals required for the calculations are linearly reducible, allowed us to
get much more easily to a representation in terms of multiple polylogarithms by integrating Feynman parameters
using the public code {\HyperInt}.
The results obtained in this way are very compact and can be evaluated in any region of the phase space
with a simple addition of a $+\ieps$ to the kinematic invariants, according to 
Feynman's prescription. For future applications, we constructed a much more efficient representation of the $\Omega_{+++}^{(0)}$ helicity amplitude in terms of classical polylogarithms up to weight three.

\section*{Acknowledgments}
We would like to thank K.~Melnikov and F.~Caola for interesting discussions and clarifications at different stages of the project and for carefully reading the manuscript.
We are particularly indebted to K.~Melnikov for having initiated this project.
M.~B.\ wishes to thank C.~Duhr for having kindly provided the package \texttt{PolyLogTools} before its official release, K.~Kudahskin for elucidating discussions about the use of {\HyperInt} in the very early stages of this project, and R.~Lee for useful discussion about differential equations and $\td\log$ forms.
V.~S.\ is grateful to C.~Duhr and R.~Lee for various pieces of advice.
Finally, we acknowledge various insightful remarks by the anonymous referee, which helped us to improve the clarity of the exposition.

M.~B.\ was supported by a graduate fellowship from the Karlsruhe Graduate School ``Collider
Physics at the highest energies and at the highest precision'' in the early stages of the project, 
and is supported by the Deutsche Forschungsgemeinschaft (DFG) under the grant no.\ 396021762 - TRR 257 
for the remaining part of it.
The work of V.~S.\ was carried out according to the research program of the Moscow Center of Fundamental and Applied Mathematics.
L.~T.\ is supported by the Royal Society through grant URF/R1/191125.

\appendix 


\section{The master integrals}
\label{sec:masters}
The following 45 planar master integrals are used as a basis for the reduction (as described in section~\ref{sec:amp-evaluation}) and as a starting point for the computation of the differential equations (see section~\ref{sec:intsDEQs}):

\begin{equation}\label{eq:reduction-basis-PL}\begin{aligned}
 &\FIPL{-1,1,1,1,1,1,1,0,0}\,, & &\FIPL{0,0,1,2,0,0,2,0,0}\,, & &\FIPL{0,0,2,2,0,0,1,0,0}\,, \\
 &\FIPL{0,1,1,0,1,1,1,0,0}\,,  & &\FIPL{0,1,1,1,0,1,1,0,0}\,, & &\FIPL{0,1,1,1,1,0,1,0,0}\,, \\
 &\FIPL{0,1,1,1,1,1,1,0,0}\,,  & &\FIPL{0,1,2,0,0,2,0,0,0}\,, & &\FIPL{0,1,2,0,1,0,1,0,0}\,, \\
 &\FIPL{0,1,2,0,2,0,0,0,0}\,,  & &\FIPL{0,1,2,1,0,0,1,0,0}\,, & &\FIPL{0,1,2,1,0,1,0,0,0}\,, \\
 &\FIPL{0,1,2,1,0,1,1,0,0}\,,  & &\FIPL{0,1,2,1,1,0,1,0,0}\,, & &\FIPL{0,1,2,1,1,1,0,0,0}\,, \\
 &\FIPL{0,2,0,2,0,1,0,0,0}\,,  & &\FIPL{0,2,0,2,0,1,1,0,0}\,, & &\FIPL{0,2,1,1,0,1,1,0,0}\,, \\
 &\FIPL{0,2,1,1,1,1,0,0,0}\,,  & &\FIPL{0,2,1,1,1,1,1,0,0}\,, & &\FIPL{0,2,2,0,0,1,0,0,0}\,, \\
 &\FIPL{0,2,2,0,1,0,0,0,0}\,,  & &\FIPL{0,2,2,0,1,0,1,0,0}\,, & &\FIPL{0,2,2,1,0,0,1,0,0}\,, \\
 &\FIPL{0,2,2,1,0,1,0,0,0}\,,  & &\FIPL{1,0,1,0,1,0,2,0,0}\,, & &\FIPL{1,0,1,1,1,0,2,0,0}\,, \\
 &\FIPL{1,0,2,0,1,0,2,0,0}\,,  & &\FIPL{1,0,2,1,1,0,1,0,0}\,, & &\FIPL{1,1,1,0,0,1,1,0,0}\,, \\
 &\FIPL{1,1,1,0,1,0,1,0,0}\,,  & &\FIPL{1,1,1,0,1,0,2,0,0}\,, & &\FIPL{1,1,1,0,1,1,1,0,0}\,, \\
 &\FIPL{1,1,1,1,0,0,1,0,0}\,,  & &\FIPL{1,1,1,1,0,1,1,0,0}\,, & &\FIPL{1,1,1,1,1,-1,1,0,0}\,,\\
 &\FIPL{1,1,1,1,1,0,1,0,0}\,,  & &\FIPL{1,1,1,1,1,0,2,0,0}\,, & &\FIPL{2,1,1,0,0,0,2,0,0}\,, \\
 &\FIPL{1,1,2,0,1,0,1,0,0}\,,  & &\FIPL{2,2,0,0,1,0,0,0,0}\,, & &\FIPL{2,2,0,0,1,0,1,0,0}\,, \\
 &\FIPL{1,2,1,0,0,0,0,0,0}\,,  & &\FIPL{2,2,0,0,0,1,0,0,0}\,, & &\FIPL{2,2,0,0,0,1,1,0,0}\,.
\end{aligned}\end{equation}
The last two master integrals, $\FIPL{2,2,0,0,0,1,0,0,0}$ and $\FIPL{2,2,0,0,0,1,1,0,0}$ do not appear in the amplitude but are required during the computation of the differential equations, since they play a role in the non-homogeneous part of the equations.

~ 

The following 18 non-planar master integrals are used as a basis for the reduction (as described in section~\ref{sec:amp-evaluation}) and as a starting point for the computation of the differential equations (see section~\ref{sec:intsDEQs}):

\begin{equation}\label{eq:reduction-basis-NP}
\begin{aligned}
 &\FINP{0,1,1,0,1,1,1,0,0}\,, & &\FINP{1,1,1,1,1,1,1,-1,-1}\,, & &\FINP{0,1,1,1,1,0,1,0,0}\,, \\
 &\FINP{1,1,0,1,1,0,1,0,0}\,, & &\FINP{0,1,1,1,1,1,1,0,-1}\,, & &\FINP{0,1,1,1,1,1,1,0,0}\,,  \\
 &\FINP{0,1,1,2,1,0,1,0,0}\,, & &\FINP{1,1,1,1,1,1,1,-1,0}\,, & &\FINP{0,1,1,2,1,1,1,0,0}\,, \\
 &\FINP{1,1,0,1,1,1,1,0,0}\,, & &\FINP{1,1,1,0,0,1,1,0,0}\,, & &\FINP{1,1,1,0,0,2,1,0,0}\,, \\
 &\FINP{1,1,1,1,0,0,1,0,0}\,, & &\FINP{1,1,1,1,1,1,1,0,-1}\,, & &\FINP{1,1,1,1,0,2,1,0,0}\,, \\
 &\FINP{1,1,1,1,0,1,1,0,0}\,, & &\FINP{1,1,1,1,0,1,1,0,-1}\,, & &\FINP{1,1,1,1,1,1,1,0,0}\,,
\end{aligned}
\end{equation}

In the basis of master integrals used in the differential equations, two more non-planar master integrals appear:
\begin{align}
		&\FINP{1,1,1,1,1,1,0,0,0}\,, & &\FINP{1,2,1,1,1,1,0,0,0}\,, & &\FINP{1,1,1,1,1,1,0,0,-1}\,.
\end{align}
These integrals can be rewritten in terms of the planar master integrals $\FIPL{1,1,1,1,1,0,1,0,0}$, $\FIPL{1,1,1,1,1,-1,1,0,0}$, and $\FIPL{1,1,1,1,1,0,2,0,0}$ and their subtopologies. We keep them as they are for simplicity, also in the ancillary files.

\section{The less divergent basis}
\label{sec:finitemasters}

The following list of finite master integrals were calculated by 
integration of their parametric representations, see section~\ref{sec:intsHyperInt}:
\begin{align}
	\label{eq:finitebasis}
	\begin{aligned}
		& \begin{aligned}
			&\mathcal{I}_{\mathsf{PL}}^{(6)}(0,3,1,1,1,1,0,0,0)\,, & &\mathcal{I}_{\mathsf{PL}}^{(6)}(0,4,1,1,1,1,0,0,0)\,, & &\mathcal{I}_{\mathsf{PL}}^{(6)}(1,1,2,0,1,0,2,0,0)\,, \\
			&\mathcal{I}_{\mathsf{PL}}^{(6)}(1,0,1,1,1,0,3,0,0)\,, & &\mathcal{I}_{\mathsf{PL}}^{(6)}(1,0,1,1,1,0,4,0,0)\,, & &\mathcal{I}_{\mathsf{PL}}^{(6)}(0,2,2,1,1,0,1,0,0)\,, \\
			&\mathcal{I}_{\mathsf{PL}}^{(6)}(0,3,2,1,1,0,1,0,0)\,, & &\mathcal{I}_{\mathsf{PL}}^{(6)}(1,1,1,1,1,0,3,0,0)\,, & &\mathcal{I}_{\mathsf{PL}}^{(6)}(1,1,1,1,1,0,2,0,0)\,, \\
			&\mathcal{I}_{\mathsf{PL}}^{(6)}(1,2,1,1,1,0,2,0,0)\,, & &\mathcal{I}_{\mathsf{PL}}^{(6)}(0,2,2,1,0,1,1,0,0)\,, & &\mathcal{I}_{\mathsf{PL}}^{(6)}(0,3,1,1,1,1,1,0,0)\,, \\
			&\mathcal{I}_{\mathsf{PL}}^{(6)}(0,2,1,1,1,1,1,0,0)\,, & &\mathcal{I}_{\mathsf{PL}}^{(6)}(0,2,1,1,1,1,2,0,0)\,, & &\mathcal{I}_{\mathsf{NP}}^{(6)}(1,1,1,3,0,0,1,0,0)\,, \\
			&\mathcal{I}_{\mathsf{NP}}^{(6)}(0,1,1,3,1,0,1,0,0)\,, & &\mathcal{I}_{\mathsf{NP}}^{(6)}(0,1,1,4,1,0,1,0,0)\,, & &\mathcal{I}_{\mathsf{NP}}^{(6)}(1,1,1,0,0,3,1,0,0)\,, \\
			&\mathcal{I}_{\mathsf{NP}}^{(6)}(1,1,1,0,0,4,1,0,0)\,, & &\mathcal{I}_{\mathsf{NP}}^{(6)}(1,1,1,2,0,1,1,0,0)\,, & &\mathcal{I}_{\mathsf{NP}}^{(6)}(1,1,1,1,0,2,1,0,0)\,, \\
			&\mathcal{I}_{\mathsf{NP}}^{(6)}(1,1,1,2,0,2,1,0,0)\,, & &\mathcal{I}_{\mathsf{NP}}^{(6)}(0,1,1,0,1,3,1,0,0)\,, & &\mathcal{I}_{\mathsf{NP}}^{(6)}(0,1,1,2,1,1,1,0,0)\,, \\
			&\mathcal{I}_{\mathsf{NP}}^{(6)}(0,1,1,1,1,2,1,0,0)\,, & &\mathcal{I}_{\mathsf{NP}}^{(6)}(0,1,1,2,1,2,1,0,0)\,, & &\mathcal{I}_{\mathsf{NP}}^{(6)}(1,1,1,2,1,1,1,0,0)\,, \\
			&\mathcal{I}_{\mathsf{NP}}^{(6)}(1,1,1,1,1,2,1,0,0)\,, & &\mathcal{I}_{\mathsf{NP}}^{(6)}(1,1,1,3,1,1,1,0,0)\,, & &\mathcal{I}_{\mathsf{NP}}^{(6)}(1,1,1,1,1,3,1,0,0)\,,	\\
			&\mathcal{I}_{\mathsf{PL}}(1,1,1,0,1,0,1,0,0)\,, & &\mathcal{I}_{\mathsf{PL}}(1,1,1,0,1,0,2,0,0)\,, & &\mathcal{I}_{\mathsf{PL}}(0,1,1,1,0,1,1,0,0)\,, \\
& \mathcal{I}_{\mathsf{PL}}(0,2,1,1,0,1,1,0,0)\,, & &\FIPL{1,0,1,0,1,0,2,0,0} \,, & &\mathcal{I}_{\mathsf{PL}}(1,1,1,0,0,1,1,0,0)\,, \\
			& & & \FIPL{1,0,1,0,1,0,2,0,0}\,, & &
		\end{aligned}	\\
	\end{aligned}
\end{align}
where the upper index $(6)$ indicates that the corresponding integral is evaluated in $d=6$ dimensions (without index, $d=4$).
The remaining master integrals below are still divergent and were integrated after regularizing integration by parts in Feynman parameters \cite{vonManteuffel:2014qoa,Panzer:DivergencesManyScales}:
\begin{align}
	\label{eq:divints}
	\begin{aligned}
			& \FIPL{1,1,1,0,1,1,1,0,0}\,, & & \FIPL{1,1,1,1,0,1,1,0,0}\,, & & \FINP{1,1,0,1,1,0,1,0,0}\,,\\
			& \FIPL{0,1,1,0,1,1,1,0,0}\,, & & \FIPL{1,1,1,1,0,0,1,0,0}\,, & &\FINP{1,1,0,1,1,1,1,0,0}\,, \\
			&\mathcal{I}_{\mathsf{PL}}(0,0,1,2,0,0,2,0,0)\,, & &\mathcal{I}_{\mathsf{PL}}(0,0,2,2,0,0,1,0,0)\,, & &\mathcal{I}_{\mathsf{PL}}(0,1,2,0,0,2,0,0,0)\,, \\
			&\mathcal{I}_{\mathsf{PL}}(0,1,2,0,1,0,1,0,0)\,, & &\mathcal{I}_{\mathsf{PL}}(0,1,2,0,2,0,0,0,0)\,, & &\mathcal{I}_{\mathsf{PL}}(0,1,2,1,0,0,1,0,0)\,, \\
			&\mathcal{I}_{\mathsf{PL}}(0,1,2,1,0,1,0,0,0)\,, & &\mathcal{I}_{\mathsf{PL}}(0,2,2,0,0,1,0,0,0)\,, & &\mathcal{I}_{\mathsf{PL}}(0,2,2,0,1,0,0,0,0)\,, \\
			&\mathcal{I}_{\mathsf{PL}}(0,2,0,2,0,1,0,0,0)\,, & &\mathcal{I}_{\mathsf{PL}}(0,2,0,2,0,1,1,0,0)\,, & &\mathcal{I}_{\mathsf{PL}}(0,2,2,0,1,0,1,0,0)\,, \\
			&\mathcal{I}_{\mathsf{PL}}(0,2,2,1,0,0,1,0,0)\,, & &\mathcal{I}_{\mathsf{PL}}(0,2,2,1,0,1,0,0,0)\,, & &\FIPL{2,2,0,0,1,0,1,0,0}\,. \\
			&\mathcal{I}_{\mathsf{PL}}(1,0,2,0,1,0,2,0,0)\,, & &\mathcal{I}_{\mathsf{PL}}(1,2,1,0,0,0,0,0,0)\,, & &\mathcal{I}_{\mathsf{PL}}(2,1,1,0,0,0,2,0,0)\,, \\
			& & & \FIPL{2,2,0,0,1,0,0,0,0}\,. & &
	\end{aligned}
\end{align}
We note that the six integrals in the top two rows of \eqref{eq:divints} have only a single pole as $d \to 4$ and they appear in the amplitude with a factor of $(d-4)$, so only the pole (leading order) of those integrals contributes to the helicity amplitudes in $d=4$.

\bibliography{Biblio}

\end{document}